\documentclass[acmsmall, nonacm]{acmart}
\AtBeginDocument{%
  }

\setcopyright{none}
\renewcommand\footnotetextcopyrightpermission[1]{}
\usepackage{svg}
\usepackage{wrapfig}
\usepackage{paralist}
\usepackage{caption}
\usepackage{subcaption}
\usepackage{float}
\usepackage{multirow} 
\usepackage{dblfloatfix}
\usepackage{pifont}
\newcommand{\cmark}{\ding{51}}  
\newcommand{\xmark}{\ding{55}}  
\newcommand{\secref}[1]{§\ref{#1}}

\usepackage[normalem]{ulem} 

\begin{document}


\title{Towards Sub-millisecond Latency and Guaranteed Bit Rates in 5G User Plane}


\author{Leonardo Alberro}
\email{lalberro@fing.edu.uy}
\affiliation{%
  \institution{University of the Republic}
  \city{Montevideo}
  \country{Uruguay}}

\author{Noura Limam}
\affiliation{%
  \institution{University of Waterloo}
  \city{Waterloo}
  \country{Canada}}

\author{Raouf Boutaba}
\affiliation{%
  \institution{University of Waterloo}
  \city{Waterloo}
  \country{Canada}}


\begin{abstract}
Next-generation services demand stringent Quality of Service (QoS) guarantees, such as per-flow bandwidth assurance, ultra-low latency, and traffic prioritization, posing significant challenges to 5G and beyond networks. As 5G network functions increasingly migrate to edge and central clouds, the transport layer becomes a critical enabler of end-to-end QoS compliance. However, traditional fixed-function infrastructure lacks the flexibility to support the diverse and dynamic QoS profiles standardized by 3GPP.

This paper presents a QoS-aware data plane model for programmable transport networks, designed to provide predictable behavior and fine-grained service differentiation. The model supports all 3GPP QoS resource types and integrates per-flow metering, classification, strict priority scheduling, and delay-aware queuing. Implemented on off-the-shelf programmable hardware using P4 and evaluated on an Intel Tofino switch, our approach ensures per-flow bandwidth guarantees, sub-millisecond delay for delay-critical traffic, and resilience under congestion. Experimental results demonstrate that the model achieves microsecond-level latencies and near-zero packet loss for mission-critical flows, validating its suitability for future QoS-sensitive applications in 5G and beyond.
\end{abstract}




\maketitle

\section{Introduction}\label{sec:intro}
Mobile networks are undergoing a paradigm shift driven by the need to accommodate diverse services with varying performance requirements, broadly categorized as enhanced Mobile Broadband (eMBB), massive Machine Type Communications (mMTC), and Ultra-Reliable Low Latency Communications (URLLC). From bandwidth-heavy real-time gaming and high-definition video streaming to massive IoT deployments and mission-critical control signaling, each service class demands specific latency, throughput, and reliability guarantees.

Network slicing in 5G networks leverages virtualization to partition a shared physical infrastructure into multiple logical networks (slices), each tailored to a particular service class and Quality of Service (QoS) profile~\cite{3gpp:slicing:TS28.530, book:slicing:kazmi2019}. By decoupling resources and configurations end-to-end across the RAN, transport, and core segments, slicing enables eMBB, mMTC, and URLLC services to coexist without cross-interference. It promises guaranteed bit rates, ultra-low delays, or massive connection capacity as required.

Recent research efforts have significantly advanced the understanding and implementation of slicing in both the core~\cite{Core68, Core46} and RAN~\cite{RAN1, RAN41, RAN15, RAN24} segments, contributing toward an end-to-end (E2E) slicing architecture for 5G.

As 5G network functions migrate toward cloud-native architectures, the transport network is increasingly critical in maintaining QoS guarantees. Traditional transport infrastructures lack the flexibility needed to support modern services' dynamic and heterogeneous requirements. Therefore, a shift toward programmable data planes has emerged as a promising solution, enabling precise and scalable traffic management without compromising throughput or latency~\cite{progammability1, progammability2, progammability3, progammability-runtime, progammability-smartNICs}. Technologies such as the P4 programming language~\cite{P4}, combined with programmable hardware platforms like Tofino switches and smartNICs~\cite{IntelTofino, netronomeSmartnics}, provide the necessary building blocks for this evolution.

In this context, a few studies have explored programmable QoS enforcement in the 5G transport (5GT), including mechanisms for per-slice bandwidth guarantees~\cite{related:P4-TINS}, delay-aware scheduling~\cite{related:virtual-queues}, and QoS-oriented data planes~\cite{related:QoS-Aware}. However, these approaches often suffer from trade-offs, such as sacrificing bandwidth for delay guarantees, and lack support for dynamic prioritization, which is mandated by 3GPP~\cite{3gpp:Priority:TS22.261}. Furthermore, they typically address slice-level requirements without accommodating the finer granularity of individual flow-level QoS.

We argue that the current state-of-the-art lacks a comprehensive QoS-aware transport data plane that fulfills next-generation services' diverse and demanding requirements. Specifically, a solution is needed that can simultaneously enforce guaranteed bit rates, deliver ultra-low latency, and dynamically prioritize traffic under network stress, all in alignment with 3GPP's standardized flow-level QoS profiles.

This paper introduces QoS-driven data plane forwarding behaviours to ensure predictable per-flow QoS in 5G transport networks. Our forwarding behaviours slice the 5G transport based on standardized QoS profiles, mapping flows to specific resource types, priority levels, and delay budgets.
We validate our design on our in-lab high-speed 5G testbed equipped with a P4-programmable Intel Tofino switch~\cite{IntelTofino}. We implement our packet handling and forwarding behaviours in the switch using a combination of P4-based techniques, such as meters, traffic shaping, packet scheduling, and multiple hardware queues, demonstrating our solution's ability to enforce per-flow bandwidth guarantees under congestion, sustain ultra-low delays for delay-sensitive traffic, and prioritize packets during network congestion.

In summary, our contributions are:
\begin{itemize}
\item A framework for QoS-aware 5GT slicing with programmable hardware, mapping 3GPP 5QI values to four distinct resource classes: Guaranteed Bit Rate (GBR), Delay-Critical Guaranteed Bit Rate (GBR*), Non-Guaranteed Bit Rate (Non-GBR), and Delay-Critical Non-Guaranteed Bit Rate (Non-GBR*).
\item A programmable data plane model with tailored packet handling and forwarding behaviours that provides per-flow bandwidth guarantees, ensures sub-millisecond latency for delay-critical traffic, and enforces packet prioritization during congestion.
\item A practical implementation and experimental evaluation using a real high-speed ASIC, supported by a custom 5G user-plane traffic generator with configurable QoS parameters.
\end{itemize}

The rest of this paper is organized as follows. Section~\secref{sec:back-related} presents background, motivation, and related work. Section~\secref{sec:qos-transportNetwork} defines the foundations for a QoS-aware 5GT. Section~\secref{sec:solution} describes our model design and analysis. Section~\secref{sec:impl-evaluation} details the implementation and evaluation results. Finally, Sections~\secref{sec:discussion} and ~\secref{sec:conclusion} discuss future works and conclude the paper.

\section{Background and Related Work}\label{sec:back-related}

This section introduces fundamental concepts underlying QoS management in 5G networks. We begin by outlining the 3GPP-defined QoS model, which provides the framework for supporting heterogeneous service requirements. We then present an overview of data plane programmability with P4, emphasizing its role in enabling fine-grained QoS enforcement. Finally, we discuss prior work on network slicing and QoS-aware forwarding, focusing on the emerging use of programmable data planes to meet the stringent demands of next-generation services.

\subsection{5G Services and QoS Requirements}

\begin{figure}[h]
  \centering
  \includegraphics[width=0.8\linewidth]{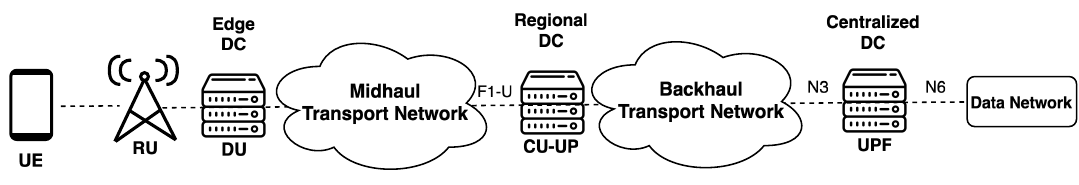}
  \caption{A 5G data plane architecture with disaggregated components placed in data centers.}
  \label{fig:oran-arch}
\end{figure}

To address the distinct needs of 5G use cases, such as eMBB's high throughput and capacity, mMTC's large number of low-data-rate connections, and URLLC extremely low latency and high reliability in support of mission-critical applications, 3GPP introduces the concept of \textit{QoS flows}, which represent the finest level of QoS granularity in the 5G ecosystem~\cite{3gpp:Priority:TS22.261}. Each QoS flow is associated with a 5G QoS Identifier (5QI) that dictates specific requirements from the network. 5QIs are mapped into a resource type, e.g., Guaranteed Bit Rate (GBR), Non-GBR, or Delay-Critical GBR, and are characterized by a set of standard (default) QoS metrics, such as flow Priority Level, end-to-end Packet Delay Budget (PDB), Packet Error Rate (PER), Averaging window (for GBR and Delay-critical GBR resource type), and Maximum Data Burst Volume (for Delay-critical GBR resource type). 5G QoS flows are also characterized by application-specific, operator-defined, or negotiated QoS parameters such as the Guaranteed Flow Bit Rate (GFBR) (for GBR resource type) and Maximum Flow Bit Rate (MFBR) (for GBR resource type). All these parameters are meant to be enforced to ensure that flows receive appropriate handling and forwarding treatment as they traverse the different segments of the 5G network. For instance, PDB defines an upper bound for the time a packet may be delayed between the UE and the N6 termination point at the User Plane Function (see Figure~\ref{fig:oran-arch}). PDB applies to the downlink packets received by the UPF over the N6 interface and/or the uplink packets sent by the UE. In the 3GPP access network, the PDB is used to support the configuration of scheduling and link layer functions (e.g., the setting of scheduling priority weights). For Delay-critical GBR QoS flows not exceeding GFBR and MDBV within the period of PDB, a packet delayed more than PDB is counted as lost. For GBR QoS Flows with GBR resource type not exceeding GFBR, 98\% of the packets shall not experience a delay exceeding PDB.

\subsection{Data Plane Programmability with P4}\label{sub:P4}

Traditional network devices offer limited flexibility in processing and managing traffic. In contrast, programmable data planes, enabled by languages such as P4, provide means to implement custom packet processing logic directly in packet forwarding hardware~\cite{P4}. This includes the ability to parse headers, match on arbitrary fields, and execute user-defined actions at line rate.

A key capability of P4 switches is the use of \textit{meters} for runtime traffic policing, traffic shaping, and rate limiting. Typically, meters implement the Two-Rate Three-Color Marker (trTCM) algorithm~\cite{trtcm-RFC2698}, which classifies packets based on two configurable thresholds: the Committed Information Rate (CIR) and the Peak Information Rate (PIR). By using two token buckets that refill at the CIR and PIR rates, the meter labels packets as green (within CIR), yellow (between CIR and PIR), or red (above PIR). This classification enables subsequent forwarding logic to enforce drop policies and apply shaping, among others.

Beyond metering, P4 programmability supports fine-grained queuing and scheduling policies that enable networks to enforce service differentiation at a granular level~\cite{related:p4-survey-traffic-policy}. These features position programmable data planes as a promising platform for realizing QoS-aware network slicing~\cite{survey-p4-5g, related:slicingP4}, as they can adapt in real time to the flow- and packet-level requirements of diverse applications.

\subsection{Programmable Data Plane for Enhanced 5G Services}

To support 5G network slicing and inter-slice performance isolation, several studies have proposed enhancements to the 5G Core (5GC)~\cite{Core68, Core46} and Radio Access Network (RAN)~\cite{RAN1, RAN41, RAN15, RAN24}, primarily focusing on control plane mechanisms for slice definition, signaling, and orchestration.

However, meeting QoS requirements requires data plane capabilities as well. With the widespread adoption of virtualization and softwarization technologies, modern mobile networks typically implement rate limiting and traffic scheduling as virtual network functions (VNFs) deployed on general-purpose CPUs~\cite{bwg9}. While VNFs offer flexibility in traffic policing, they increase delay and jitter, which are particularly detrimental to delay-sensitive applications~\cite{bw-slicing}.

To address these limitations, recent efforts have turned to data plane programmability, especially using the P4 language, to enforce QoS guarantees directly in the forwarding path~\cite{related:QoS-Aware, related:dataplaneprogrammability, related:P4-TINS}. Unlike fixed-function switches, P4-programmable devices allow fine-grained traffic control at line rate.

A few works have explored how programmable data planes can be used for 5G slicing. For instance,~\cite{related:virtual-queues} introduced virtual queues within the P4 pipeline to manage traffic and meet latency constraints. However, this design does not ensure bandwidth guarantees or eliminate inter-flow interference. In~\cite{related:P4-TINS}, the authors used P4 meters and priority scheduling to allocate bandwidth per slice but did not address the needs of delay-sensitive flows or QoS-driven flow prioritization. The work in~\cite{related:TCP-friendly} proposed a P4-based scheme tailored to TCP traffic, aiming to reduce performance degradation due to metering behavior. Meanwhile,~\cite{related:slicingP4} surveyed various slicing strategies and implemented them on both software and hardware P4 targets, with a primary focus on achieving slice isolation rather than comprehensive QoS enforcement.

Beyond slicing, several studies have advanced specific functions within programmable data planes, particularly around packet scheduling. Works such as~\cite{related:sifter, pFabric, pscheduler2, related:DRPIFO} have introduced schedulers to ensure fair bandwidth allocation, reduce flow completion times, or provide bounded delay. However, these proposals typically tackle isolated challenges and do not offer a unified solution capable of enforcing QoS across diverse traffic types.

In contrast, our approach takes a holistic view: it leverages the full capabilities of P4 to implement a QoS-aware transport data plane that supports per-flow behaviour in line with 3GPP’s standardized 5QI profiles. Our design enables simultaneous enforcement of bandwidth guarantees, delay predictability, and prioritization, offering a unified and programmable foundation for next-generation 5G services. 
Table~\ref{tab:related-work} summarizes the comparison between this work and related approaches. Although existing solutions address most of the requirements individually, they exhibit fundamental limitations that hinder the integration of multiple techniques without trade-offs (e.g., achieving delay-critical support at the expense of per-flow bandwidth guarantees). In contrast, the proposed solution considers all the requirements from the design and can accurately model realistic and standardized 5G QoS profiles.

\begin{table}[t]
\caption{Comparison of related work on QoS enforcement and 5G network slicing using programmable data planes}
\label{tab:related-work}
\scalebox{0.75}{
\begin{tabular}{@{}lccccc@{}}
\toprule
\multicolumn{1}{c}{Work}                        & \begin{tabular}[c]{@{}c@{}}QoS Profiles \\ Supported\end{tabular} & \begin{tabular}[c]{@{}c@{}}Per-Flow\\ Granularity\end{tabular} & \begin{tabular}[c]{@{}c@{}}Delay-Critical \\ Support\end{tabular} & \begin{tabular}[c]{@{}c@{}}Bandwidth\\ Guarantee\end{tabular} & \begin{tabular}[c]{@{}c@{}}Prioritization \\ Support\end{tabular} \\ \midrule
P4-TINS~\cite{related:P4-TINS}                  & Limited (per-slice)                                               & \xmark                                                         & \xmark                                                            & \cmark                                                        & \xmark                                                            \\
P4-NetFPGA~\cite{related:QoS-Aware}             & Custom                                                            & \cmark                                                         & \xmark                                                            & \xmark                                                        & \cmark                                                            \\
P4 Virtual Queues~\cite{related:virtual-queues} & Limited (per-slice)                                               & \xmark                                                         & \cmark                                                            & \xmark                                                        & \xmark                                                            \\
Slicing with P4~\cite{related:slicingP4}        & None                                                              & \xmark                                                         & \xmark                                                            & \xmark                                                        & \xmark                                                            \\
P4-TCP~\cite{related:TCP-friendly}              & None                                                              & \cmark                                                         & \xmark                                                            & \xmark                                                        & \xmark                                                            \\
This work                                       & Full (3GPP 5QI)                                                   & \cmark                                                         & \cmark                                                            & \cmark                                                        & \cmark                                                            \\ \bottomrule
\end{tabular}
}
\end{table}

\begin{table}[ht]
  \caption{Examples of standardized 5QI to QoS characteristics mapping. The PDB value includes the CN PDB value. Services with CN PDB < 10 ms may be assumed to be located "close" to the RAN. }
  \label{tab:5qis}
  \centering
  \scalebox{0.75}{
\begin{tabular}{@{}ccccccc@{}}
\toprule
5QI & \begin{tabular}[c]{@{}c@{}}Resource\\ Type\end{tabular} & \begin{tabular}[c]{@{}c@{}}Default\\ Priority\end{tabular} & \begin{tabular}[c]{@{}c@{}}PDB\\ (ms)\end{tabular} & \begin{tabular}[c]{@{}c@{}}CN PDB\\ (ms)\end{tabular} & PER       & Example Services                                  \\ \midrule
2   & GBR                                                     & 40                                                         & 150                                                & 20                                                    & $10^{-3}$ & Conversational Video (Live Streaming)             \\
4   &                                                         & 50                                                         & 300                                                & 20                                                    & $10^{-3}$ & Non-Conversational Video (Buffered Streaming)     \\
65  &                                                         & 7                                                          & 75                                                 & 10                                                    & $10^{-2}$ & Mission Critical user plane Push To Talk voice    \\
67  &                                                         & 15                                                         & 100                                                & 20                                                    & $10^{-3}$ & Mission Critical Video user plane                 \\
7   & Non-GBR                                                 & 70                                                         & 100                                                & 20                                                    & $10^{-3}$ & Voice, Video (Live Streaming), Interactive Gaming \\
69  &                                                         & 5                                                          & 60                                                 & 10                                                    & $10^{-6}$ & Mission Critical delay sensitive signalling       \\
80  &                                                         & 68                                                         & 10                                                 & 2                                                     & $10^{-6}$ & Low Latency eMBB applications Augmented Reality   \\
84  & GBR*                                                    & 24                                                         & 30                                                 & 5                                                     & $10^{-5}$ & Intelligent transport systems                     \\
86  &                                                         & 18                                                         & 5                                                  & 2                                                     & $10^{-4}$ & V2X messages (Advanced Driving)                   \\
89  &                                                         & 25                                                         & 15                                                 & 1                                                     & $10^{-4}$ & Visual content for cloud/edge/split rendering     \\ \bottomrule
\end{tabular}
}
\end{table}


\section{A QoS-aware Transport Network} \label{sec:qos-transportNetwork} 

Classifying 5G slices into the three service \emph{archetypes} eMBB, mMTC, and URLLC brings practical benefits with regard to resource provisioning and network dimensioning. Each category carries distinct resource demands; eMBB needs wide bandwidth and multi-gigabit backhaul, mMTC calls for support of millions of low-rate connections, and URLLC requires finely-tuned scheduling to meet sub-millisecond delays. By classifying traffic into these categories, operators can dimension RAN parameters (e.g., scheduling algorithms), transport links (e.g., slice bandwidth reservations), and core functions (e.g., edge compute placement) in a targeted way, rather than over-provisioning for the worst case of every application. 

However, this tripartite classification does not capture the fine‐grained QoS requirements of particular flows. For example, within the URLLC category, mission-critical control and V2X services differ in both their guaranteed bit-rate requirements and delay budgets, and the so-called Low-Latency eMBB applications (5QI 80) demand URLLC-style latencies despite the eMBB label (Table~\ref{tab:5qis}).

To address this gap, we propose leveraging standardized 5QI profiles for QoS-aware slicing in the 5GT. By carrying each flow’s 5QI through the transport network, we can map it directly to its precise QoS parameters—and enforce them in hardware. 

\begin{figure}[h]
  \centering
  \includegraphics[width=0.6\linewidth]{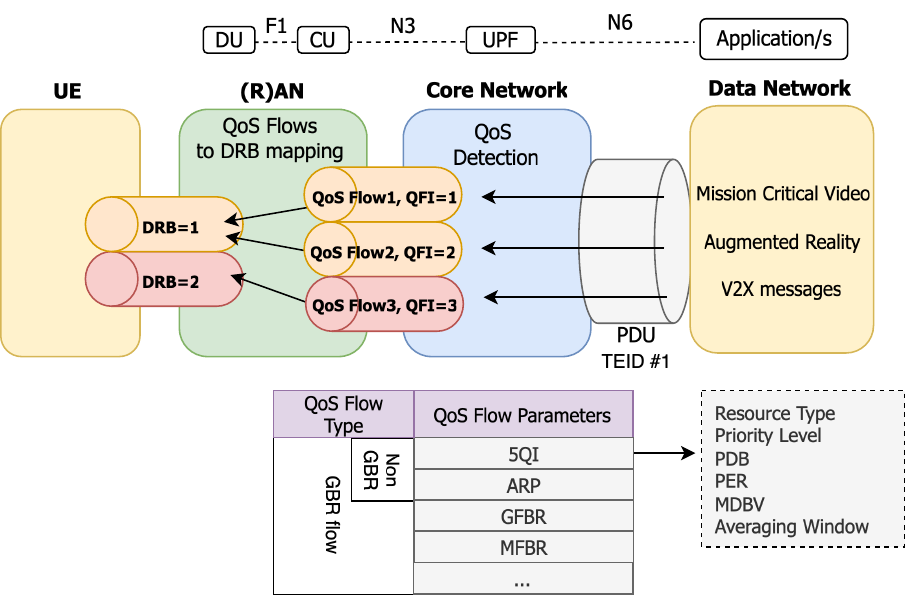}
  \caption{3GPP QoS realization for downlink packets.}
  \label{fig:qos-arch}
\end{figure}

Figure~\ref{fig:qos-arch} illustrates the standardized QoS treatment across the network. On the downlink, QoS detection is enforced in the CN, and flows are categorized into different QoS flows and marked accordingly. Then, the QoS flows are mapped to Data Radio Bearers (DRB) channels that carry user data between the RAN and the UE.

A closer look at Table~\ref{tab:5qis} reveals non-GBR services, most notably 5QI 80, that nonetheless require ultra-low delays in the core network. To encapsulate these mixed demands, we introduce a fourth resource type, Delay-Critical Non-GBR (Non-GBR*). Our enhanced 5QI framework thus covers four resource types: GBR, Delay-Critical GBR (GBR*), Non-GBR, and Delay-Critical Non-GBR (Non-GBR*).

Finally, we must decide where and how to convey 5QI metadata to the 5GT. User traffic in 5G is encapsulated in GTP-U tunnels, identified by a 4-byte Tunnel Endpoint Identifier (TEID), over UDP source port 2152 between the DU (via F1-U) and the UPF. In our proof-of-concept implementation, we use different UDP source port ranges for different 5QIs and a mapping table shared between the core and the transport control planes, binding each 5QI to a unique range of source port numbers. Mapping 3GPP slices to transport network slices using source port numbers is also suggested in~\cite{ietf-dmm-tn-aware-mobility, ietf-teas-5g-network-slice-application}. Indeed, there are other ways to do the mapping. With modern P4‐programmable switches able to parse headers across multiple layers, any solution that embeds QoS context without altering the GTP-U header format, maintaining full compatibility with 5G standard interfaces, is technically possible. Now, when it comes to \textit{where} in the transport the QoS should be enforced, Figure~\ref{fig:qos-arch} shows that QoS information is embedded within the midhaul and the backhaul segments. As earlier enforcement is generally preferable, the transport network can apply QoS policies immediately after the F1-U interface (i.e., in the midhaul transport segment) for uplink traffic and after the N3 interface (i.e., in the backhaul transport segment) for downlink traffic.

In summary, we propose a QoS-aware data plane model for programmable hardware to ensure predictable behavior in midhaul and backhaul transport networks. Our model supports all 3GPP standardized QoS profiles, providing per-flow guaranteed bit rates, sub-millisecond latency, and prioritization for mission-critical traffic during congestion. In particular, the current design and implementation integrate and account for the following QoS parameters:
\begin{itemize}
  \item \textbf{Resource type} (i.e., GBR, Delay-Critical GBR, Non-GBR, or \textit{Delay-Critical Non-GBR}), which governs bandwidth guarantees and packet scheduling;
  \item \textbf{Priority level (or ARP)} for packet scheduling and congestion management; 
  \item \textbf{GFBR} and \textbf{MFBR} values, calculated over \textbf{Averaging window} duration, which translate into the CIR and PIR settings of P4 meters for precise rate control; and
  \item \textbf{Maximum Data Burst Volume (MDBV)}, which is used to dimension the Committed Burst Size (CBS) and PBS (Peak Burst Size) settings of P4 meters for absorbing traffic bursts.
\end{itemize}



\section{Programmable Data Plane Model}\label{sec:solution}

This section presents our programmable data plane model, designed to ensure predictable latency and throughput guarantees for specific traffic flows in 5G networks.

The proposed model supports the QoS requirements of next-generation applications by integrating per-flow rate limiting, packet classification, scheduling, and prioritization mechanisms. For flows characterized by GBR and GBR* resource type, the model provides per-flow bandwidth guarantees, improving end-user experience. Additionally, it offers predictable sub-millisecond per-packet delay for GBR* and Non-GBR* flows, even under conditions of extreme congestion. Best-effort service is also supported for Non-GBR flows or GBR flows that exceed their committed rates. High-priority flows can be prioritized without degrading the guarantees provided to other service classes.

Our design is compatible with standard programmable forwarding mechanisms, relying on match-action pipelines and widely supported P4 constructs and externs (e.g., registers, meters). It is intended to be implemented on commodity programmable switches that support hierarchical queues and basic scheduling logic.

\subsection{Design Overview}\label{sub:design}

The packet processing pipeline is organized into three main stages:
\begin{inparaenum}[1)]
\item classification,
\item metering and policing, and
\item scheduling.
\end{inparaenum}

\begin{figure}[h]
\centering
\includegraphics[width=0.95\linewidth]{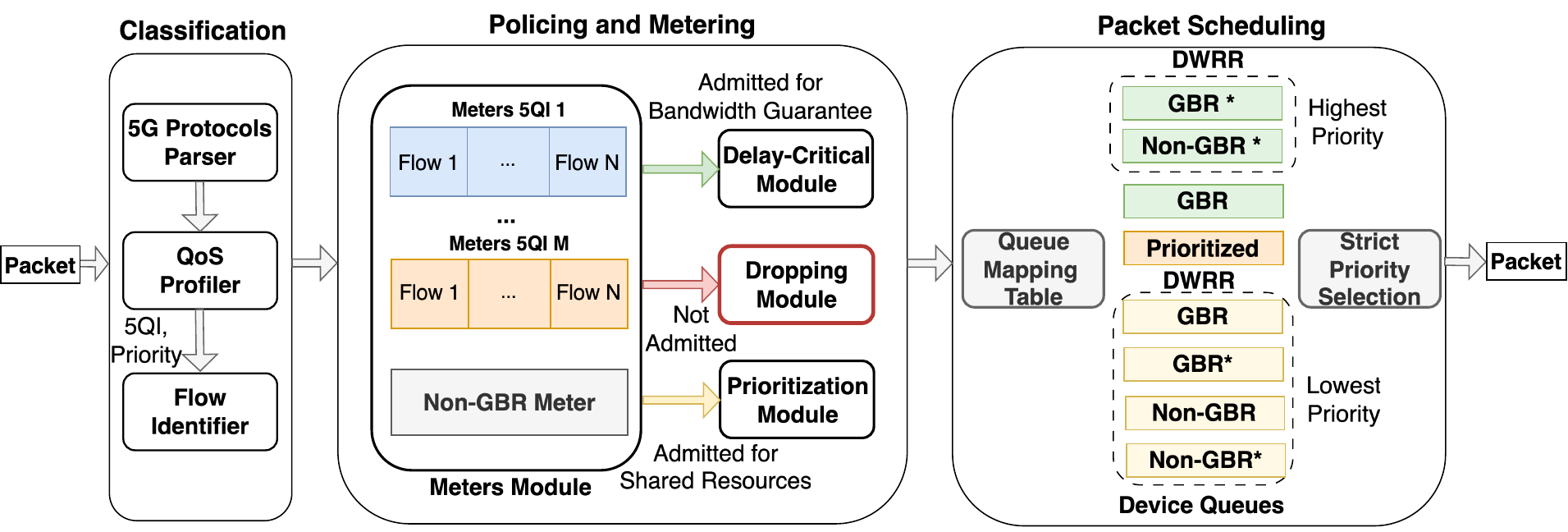}
\caption{Programmable pipeline for per-packet QoS processing.}
\label{fig:packet-flow}
\end{figure}

Figure~\ref{fig:packet-flow} illustrates the logical flow of a packet through these stages. Upon arrival, packets are classified according to their QoS profile and flow attributes. They are then subject to metering and policing to determine eligibility for bandwidth guarantees, prioritization, or best-effort treatment. Finally, packets are queued and scheduled for transmission based on their service tags and hardware capabilities.

To facilitate modularity and adaptability across different platforms, classification and marking modules store tags in a temporary metadata header field within the packet, enabling efficient processing throughout the pipeline.

\subsubsection{Classification}

This stage begins with a parser designed to extract the full 5G user-plane header stack. It supports dissection of Layer 2 Ethernet/VLAN frames and Layer 3–4 headers, including GTP-U. This enables compatibility with packets traversing between the DU in the RAN and the UPF in the core network.

Next, the QoS Profiler uses a match-action table populated by the control plane to map flow identifiers (e.g., inner UDP source port) to QoS parameters such as 5QI, resource type, and priority. These attributes are then encoded into temporary metadata.

Finally, the Flow Identifier assigns each packet to a flow. Unlike generic hash-based identification—which is susceptible to collisions~\cite{hash-collision}—we rely on the TEID field from the GTP-U header. This allows control-plane-driven flow management using a collision-free TEID space, aligned with the needs of fine-grained per-flow treatment.

\subsubsection{Metering and Policing}

Following classification, packets are processed by a matrix of traffic meters. For GBR and GBR* flows, the two-rate three-color marker (trTCM) algorithm is employed~\cite{trtcm-RFC2698}, which classifies packets as green, yellow, or red based on conformance to CIR and PIR. Green packets are eligible for guaranteed resources, yellow packets are considered for shared resources, and red packets are dropped.

Since QoS profiles are specified per 5QI, but metering is enforced per flow, we maintain a two-dimensional matrix of meters indexed by 5QI and flow ID. For Non-GBR 5QIs, we apply single-rate two-color meters (srTCM)~\cite{srtcm-RFC2697} to enforce aggregate rate limits with basic drop policies.

After metering, delay-sensitive packets (GBR* and Non-GBR*) are tagged as \textit{Delay-Critical}, while GBR flows retain the \textit{Dedicated} label. Yellow packets, exceeding their CIR, are further processed by the prioritization module. This module uses a configurable threshold on the standardized 5QI priority to identify mission-critical traffic. Packets above the threshold are marked as \textit{Prioritized}; others as \textit{Shared}.

\subsubsection{Scheduling}

The scheduling stage manages packet enqueuing and transmission. Our design leverages strict priority scheduling across multiple queues, supported by common switch hardware.

To provide isolation and fairness, we implement per-resource-type queue mapping. Instead of mapping each 5QI to a unique queue, flows are grouped based on four resource types, balancing queue utilization and hardware constraints.

Queue selection depends on both the packet’s service tag (e.g., \textit{Delay-Critical}, \textit{Shared}) and its resource type. We define two priority tiers: high (for \textit{Dedicated} and \textit{Delay-Critical}) and low (for \textit{Prioritized} and \textit{Shared}). Within each tier, the resource type determines the specific queue. Deficit Weighted Round-Robin (DWRR)~\cite{paper:DWRR} ensures fair bandwidth sharing among queues of the same priority.

Strict priority scheduling offers simplicity and portability, but can starve lower-priority queues under heavy load. To mitigate this, the control plane configures minimum bandwidth reservations for shared queues.

\subsubsection{Control Plane Integration}

The control plane oversees the configuration and monitoring of the programmable pipeline. Its responsibilities include:

\begin{itemize}
\item \textbf{Classification:} Maintaining QoS mapping tables linking UDP ports to 5QI, resource type, and priority.
\item \textbf{Policing:} Initializing and updating meters with appropriate CIR/PIR/CBS/PBS values per flow and 5QI.
\item \textbf{Scheduling:} Configuring physical queues and managing queue mapping tables for service tags and resource types.
\item \textbf{Monitoring:} Collecting performance counters and registers (e.g., delay and drop statistics) and triggering reconfiguration if QoS objectives are violated.
\end{itemize}

The control plane also manages the forwarding tables. These tables determine the output port for a packet, where the Packet Scheduler module operates.
 
\subsection{On the predictability of delay and throughput} \label{sub:PHB}
Having established the design of a programmable device based on our data plane model, we now proceed to formalize the conditions required to guarantee predictable throughput and delay. For the sake of clarity of exposition, we simplify the queueing model by assuming a single queue per priority level, resulting in a four-priority system rather than distinct queues per resource type.

Let \( S \) denote the set of all standardized 5QI identifiers, and let \( S_{\text{G}} \subset S \) represent the subset of 5QIs associated with resource types other than \textit{Non-GBR}. 
For each \( s \in S_G \), let \( F_s \) be the set of active flows mapped to \( s \). To support bandwidth guarantees, each flow \( f \in F_s \) is associated with a CIR, denoted \( \text{CIR}^f_s \), and a PIR, denoted \( \text{PIR}^f_s \). Non-GBR flows have an aggregated $PIR_{NG}$.  All traffic is assumed to share a single output link of capacity \( R \).

To ensure that the bandwidth guarantees (CIRs) can be met without contention, the following condition must hold:

\begin{displaymath} \label{eq:cir-R-equal}
  \sum_{s\in S_G}\sum_{f\in F_s}CIR^f_s \leq R
\end{displaymath}

The left-hand side of Equation~\eqref{eq:cir-R-equal} represents the aggregate minimum rate required to satisfy the committed bandwidth guarantees for flows associated with higher-priority queues, those allocated dedicated resources. 

Given the use of strict priority scheduling, the residual bandwidth, i.e., the capacity not consumed by guaranteed flows, is available to serve lower-priority queues (shared resources). We denote this residual rate by $\Delta r$, defined as:

\begin{equation}
  \Delta r = R - R_h
\end{equation}

$\Delta r$ determines the service rate available to lower-priority traffic after satisfying per-flow bit rate guarantees. It plays a critical role in bounding the delay of non-guaranteed traffic and in preventing starvation under persistent high-priority load.

To avoid starvation and ensure that shared queues receive a non-zero portion of the bandwidth, we impose the following condition:

\begin{equation} \label{eq:sum_CIRs}
  \Delta r > 0 \rightarrow \sum_{s\in S_G}\sum_{f\in F_s}CIR^f_s < R
\end{equation}

In the same manner we defined $R_h$, we can determine the maximum available rate for the medium priority queue as:


\begin{equation}
  R_m = min(\Delta r, PIR_{NG} \times p + \sum_{s\in S_G}\sum_{f\in FP_s} PIR^f_s - CIR^f_s), FP_s\subseteq F_s
\end{equation}

Here, \( FP_s \subseteq F_s \) is the subset of prioritized flows in class \( s \), and \( p \in [0,1] \) is the fraction of Non-GBR capacity allocated to prioritized traffic. 
Finally, the maximum available rate for the rest of the shared resources (lowest priority) can be simplified as the remaining bandwidth:

\begin{equation}
  R_l = \Delta r-R_m 
\end{equation}

This formulation provides a foundation for determining the maximum number of flows requiring dedicated resources that can be admitted while still ensuring per-flow bandwidth guarantees. It also establishes how the remaining (shared) bandwidth is allocated among prioritized and best-effort flows.

Under a strict priority scheduling scheme, lower-priority queues are served only after all higher-priority queues have been emptied. To prevent starvation of the shared-resource queues, we enforce the constraint in Equation~\eqref{eq:sum_CIRs}, meaning $R_h < R$.
$R_h$ also represents the maximum aggregated rate of packets going to higher priority queues. 

Let us now analyze the worst-case queueing delay at each priority level. While we conceptually partition the output capacity into $R_h$, $R_m$, and $R_l$, in practice, all packets, regardless of priority, are transmitted over the shared physical link of capacity $R$. Thus, the queueing delay depends on how the priority scheduler serializes access to this shared resource.

At the highest priority level (serving delay-critical traffic), the worst-case scenario occurs when packets arrive at the sustained rate $R_h$. Since the traffic arrival rate is lower than the service rate ($R_h < R$), the system is stable. The priority scheduler serves this queue first, meaning that, assuming discrete arrivals, there is no queueing delay for those packets. Then, the maximum queueing delay will depend on possible bursts and the buffer size for the queue.

If we assume a burst fills a buffer of size $B$, the maximum queueing delay is the time required to empty the entire buffer:

\begin{displaymath}
  D_h = \frac{B}{R}
\end{displaymath}

At the next-highest priority level (serving guaranteed but non-delay-sensitive traffic), the worst-case delay occurs when another full buffer at this level follows a full buffer of higher-priority packets. Thus, the maximum queueing delay is:

\begin{displaymath}
  D_s = \frac{2B}{R}
\end{displaymath}

In the remaining queues (shared resources), the arrival rates can exceed the portion of throughput available ($R_m + R_l$).
These arrival rates can be calculated as follows:

\begin{displaymath}
  R_{in_M} = PIR_{NG} \times p + \sum_{s\in S_G}\sum_{f\in FP_s} PIR^f_s - CIR^f_s, FP_s\subseteq F_s
\end{displaymath}

\begin{displaymath}
  R_{in_L} = PIR_{NG} \times (1-p) + \sum_{s\in S_G}\sum_{f\in FR_s} PIR^f_s - CIR^f_s, FR_s=F_s-FP_s
\end{displaymath}

Here $R_{in_M}$ is the maximum arrival rate for the medium priority queue and $R_{in_L}$ for the lowest.

Since we cannot guarantee that $R_{in_M} < Rm$ or that $R_{in_M}+R_{in_L} \leq \Delta r$, queueing delays are unavoidable in the worst case, and the maximum delays of the medium priority queue ($D_m$) and lowest priority queue ($D_l$) can be approximated by:

\begin{displaymath}
  D_m = \frac{B}{R_m}
\end{displaymath}

\begin{displaymath}
  D_l = \frac{B}{R_l}
\end{displaymath}

Notably, the condition $\Delta r > 0$ ensures that $R_m > 0$, while $R_l$ can be zero, indicating that the lowest queue may experience starvation. While this may seem problematic, it reflects a deliberate prioritization: the medium-priority queue is dynamically adjusted to serve urgent traffic bursts, whereas the lowest-priority queue serves elastic or background flows. The control plane can mitigate starvation by adjusting $R_h$ to increase $\Delta r$, or by explicitly limiting the input rate of Non-GBR traffic to bound $R_{in_L}$.

\section{Implementation and Evaluation}\label{sec:impl-evaluation}

\begin{figure}[ht]
  \centering
  \includegraphics[width=0.8\linewidth]{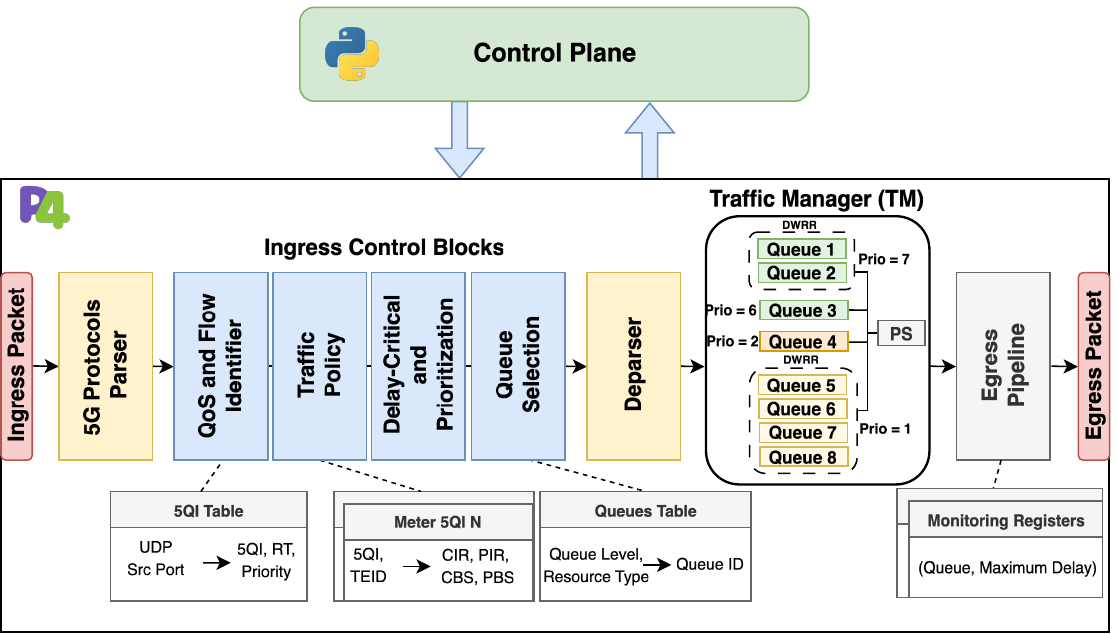}
  \caption{Implemented pipeline on the Tofino Native Architecture.} 
  \label{fig:implementation}
\end{figure}

We implemented the proposed programmable data plane model using the P4 language on an Intel Tofino programmable switch ASIC~\footnote{All source code and implementation artifacts will be made available during submission.~\cite{code-repo}}. Figure~\ref{fig:implementation} illustrates the implementation architecture. As shown, all processing stages—except for the packet scheduling logic—are realized within the ingress pipeline of the Tofino Native Architecture (TNA)~\cite{IntelTNA}. Packet scheduling relies on a combination of a programmable queue selection module and a configurable hardware component managed by the Traffic Manager (TM) within the TNA.


In this platform, meters are implemented as arrays, with a maximum size of 24,576 entries. To support the meter matrix defined in our data plane, we allocate one array per standardized 5QI with resource types other than Non-GBR. Each of the 23 arrays is defined to support up to 8K flow-specific meters. For Non-GBR 5QIs, a single aggregate meter is instantiated to enforce global rate control.

The packet scheduler leverages the target’s support for both logical and physical queues. From the data plane perspective, 32 logical queues are available. These queues can be mapped to physical output queues based on port configuration within the TM. Specifically, each 100G port group supports up to 32 queues, which can be subdivided into four 10G groups with up to eight queues each. To enforce traffic isolation and implement our priority-based scheduling strategy, we carefully configure the mapping between logical queue identifiers (used in the P4 pipeline) and their corresponding TM-configured physical queues.

Another noteworthy implementation detail concerns the monitoring of per-packet delay. Since the egress pipeline in the TNA lacks access to metadata computed during ingress, we introduce a custom bridged header at the ingress stage. This header carries the packet’s ingress timestamp and the assigned queue identifier. The information is retrieved at the egress stage to compute both processing and queueing delays, after which the header is stripped prior to transmission.

For the control plane, we developed a Python-based controller interfacing with the Intel BFRT (Barefoot Runtime) gRPC server. The controller implements all functionalities described in Section~\ref{sec:solution}, including classification rule management, meter configuration, queue mapping, and runtime performance monitoring.

\subsection{Evaluation}
\textbf{Testbed Setup.} We deployed our programmable data plane in a real-world testbed comprising an Intel Tofino switch connected to three servers via 10 Gbps links, along with a Multi-access Edge Computing (MEC) node operating at 10/25 Gbps. The control plane runs on a compute node co-located with the Tofino switch and communicates over a 10 Gbps virtual link.
\\
\textbf{Network Load.} To evaluate performance under varying conditions, we generate traffic consisting of 1500-byte GTP-U packets directed toward the MEC node. The total load ranges from 90\% to 110\% of the switch–MEC link capacity, while the aggregate committed bit rate varies between 50\% and 75\%. For the number of concurrent flows, we adopt the traffic load categories defined in~\cite{related:example-number-flows}, which reflect the operational limits of our target hardware. Accordingly, we classify load into three levels: low (up to 1,000 flows), medium (up to 2,000 flows), and high (up to 4,000 flows).
\\
\textbf{Traffic.} We employ a synthetic 5G traffic generator to emulate realistic packet behavior.
A custom generator was necessary due to the lack of existing 5G packet traces including full QoS features.
For the traffic, four representative 5QIs are selected, each corresponding to a different resource type. For simplicity, we refer to them by their associated resource class: Non-GBR, Non-GBR*, GBR, and GBR*.
\\
\textbf{Meters Setup.} For each 5QI, all associated meters are configured uniformly with the same CIR, PIR, CBS, and PBS. Each active flow is dynamically mapped to a unique meter instance with these parameters. For Non-GBR 5QIs, a single shared meter is used for all flows. In all configurations, the CIR is set to the Guaranteed Flow Bit Rate (GFBR), and the PIR to the Maximum Flow Bit Rate (MFBR). For Non-GBR flows, the PIR is configured to match their aggregate sending rate.
\\
\textbf{Metrics.} We evaluate our implementation using three key performance metrics: throughput, delay, and packet loss. Delay measurements encompass both packet processing and queueing delays, capturing the most significant sources of variability within the switch. Unless stated otherwise, all reported results represent averages across ten independent test runs.

\subsubsection{Functional evaluation at scale}\label{sub:funcEval}

To validate our model, we performed a series of tests with aggregated sending meeting 100\% of the link capacity. Under this configuration, no additional capacity is available to accommodate traffic bursts, making the system highly susceptible to congestion and packet loss due to fluctuations in packet arrivals. Generated flows are evenly distributed among the 5QIs and transmit at their MFBR, which is twice their GFBR. The per-flow sending rates for the low, medium, and high load scenarios are 10,000 kbps, 5,000 kbps, and 2,500 kbps, respectively.

\begin{figure*}[ht]
    \begin{subfigure}{0.30\linewidth}
        \centering
        \includegraphics[width=\linewidth]{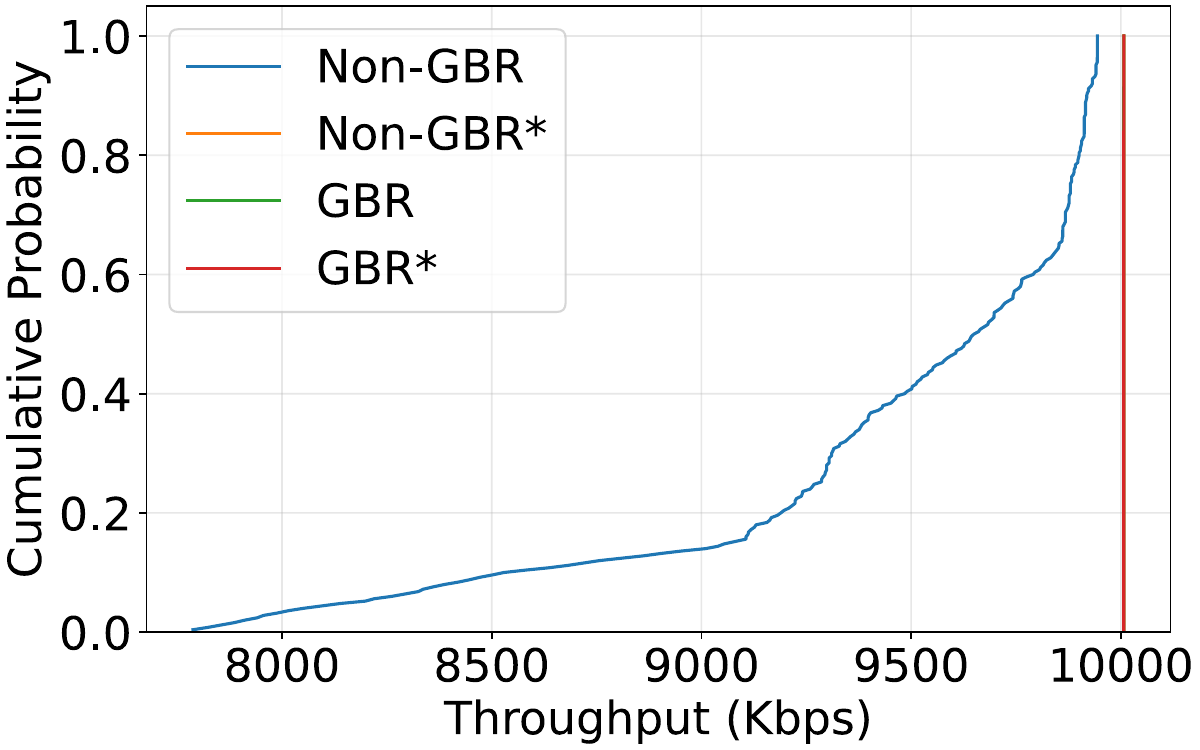}
        \caption{Low load}
        \label{fig:cdf-thr-1k}
    \end{subfigure}
    \begin{subfigure}{0.30\linewidth}
        \centering
        \includegraphics[width=\linewidth]{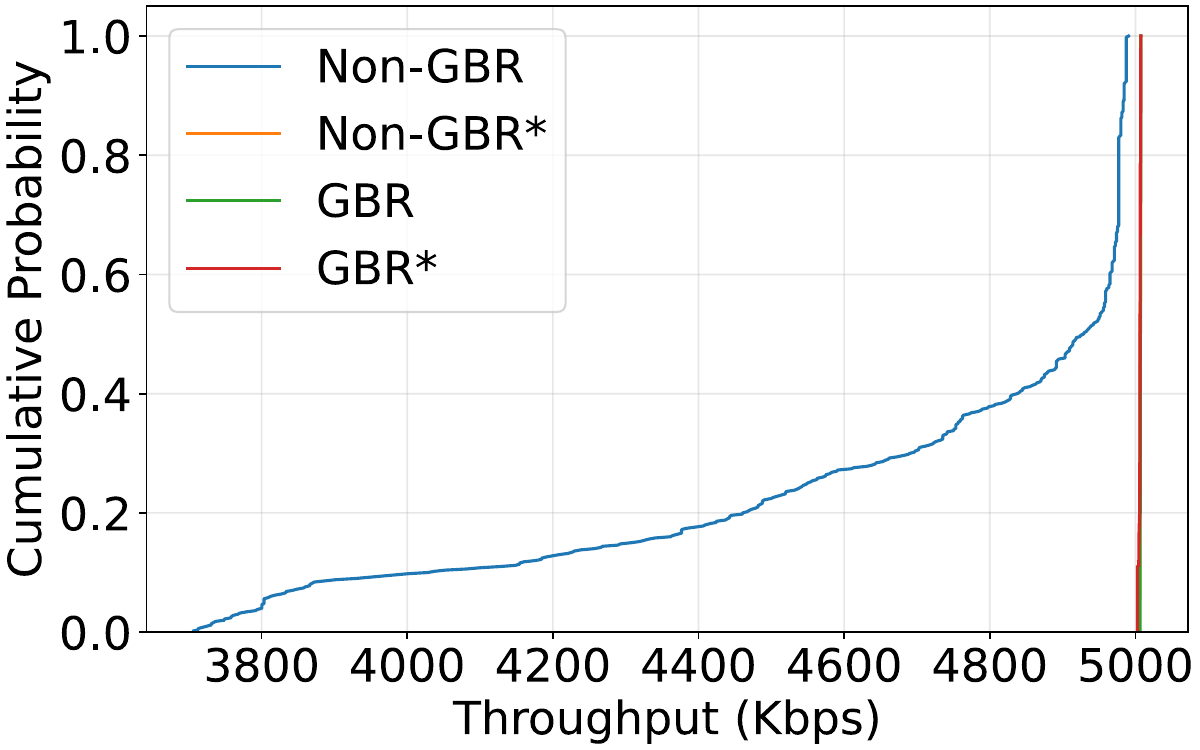}
        \caption{Medium load}
        \label{fig:cdf-thr-2k}
    \end{subfigure}
    \begin{subfigure}{0.30\linewidth}
        \centering
        \includegraphics[width=\linewidth]{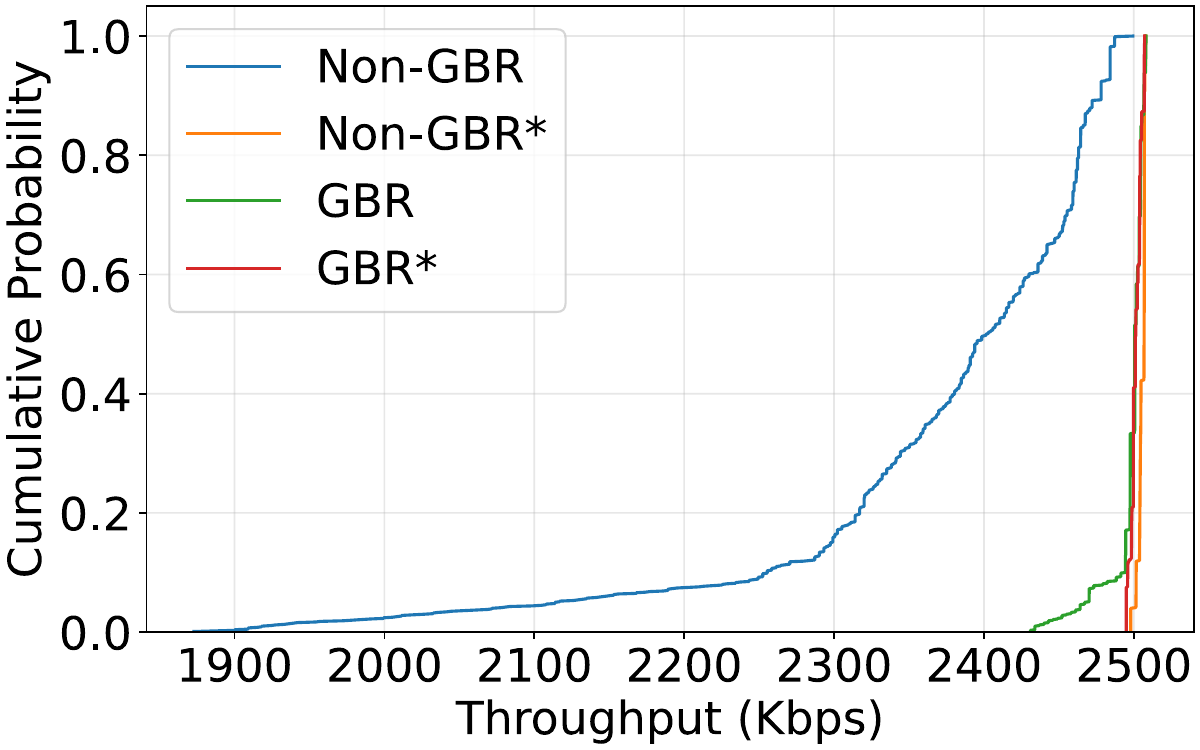}
        \caption{High load}
        \label{fig:cdf-thr-4k}
    \end{subfigure}

     \begin{subfigure}{0.30\linewidth}
        \centering
        \includegraphics[width=\linewidth]{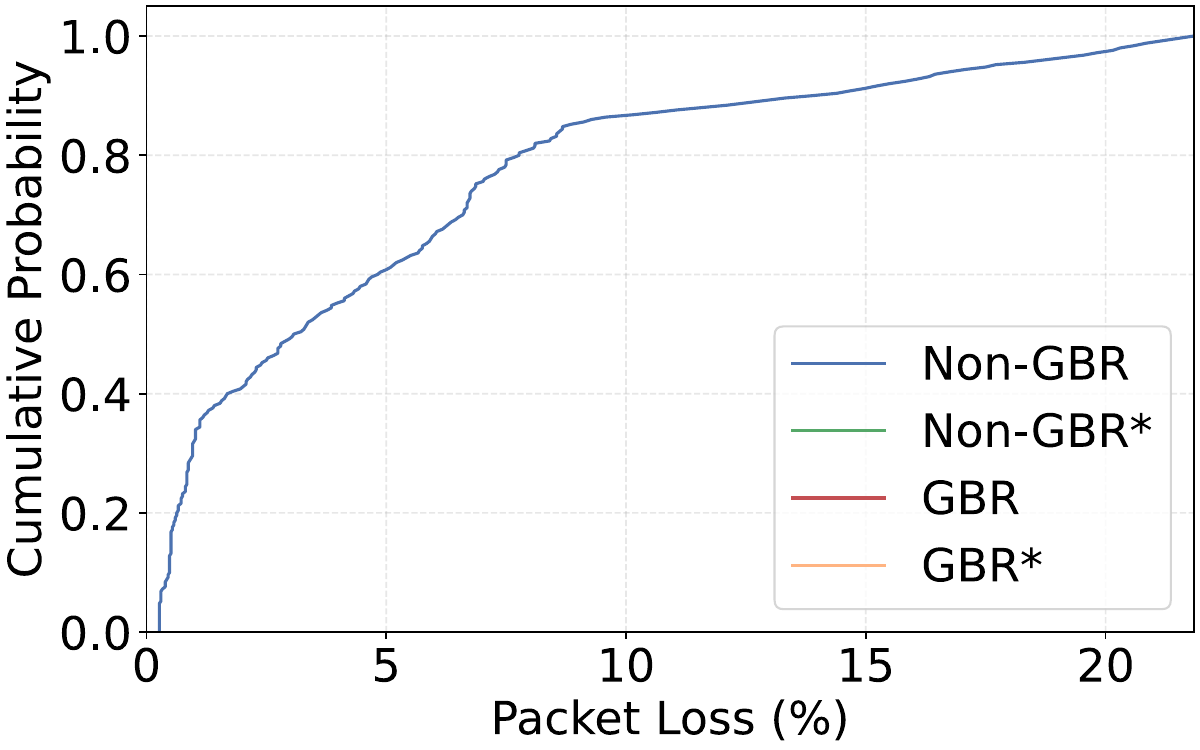}
        \caption{Low load}
        \label{fig:cdf-loss-1k}
    \end{subfigure}
    \begin{subfigure}{0.30\linewidth}
        \centering
        \includegraphics[width=\linewidth]{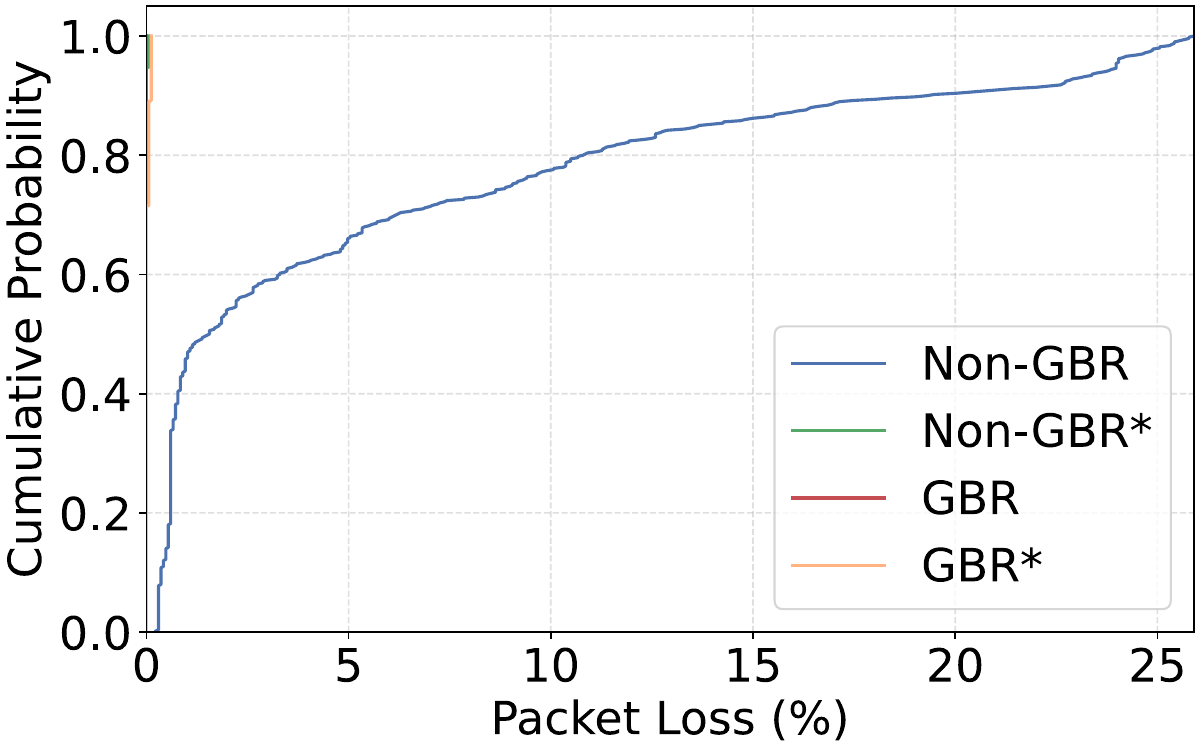}
        \caption{Medium load}
        \label{fig:cdf-loss-2k}
    \end{subfigure}
    \begin{subfigure}{0.30\linewidth}
        \centering
        \includegraphics[width=\linewidth]{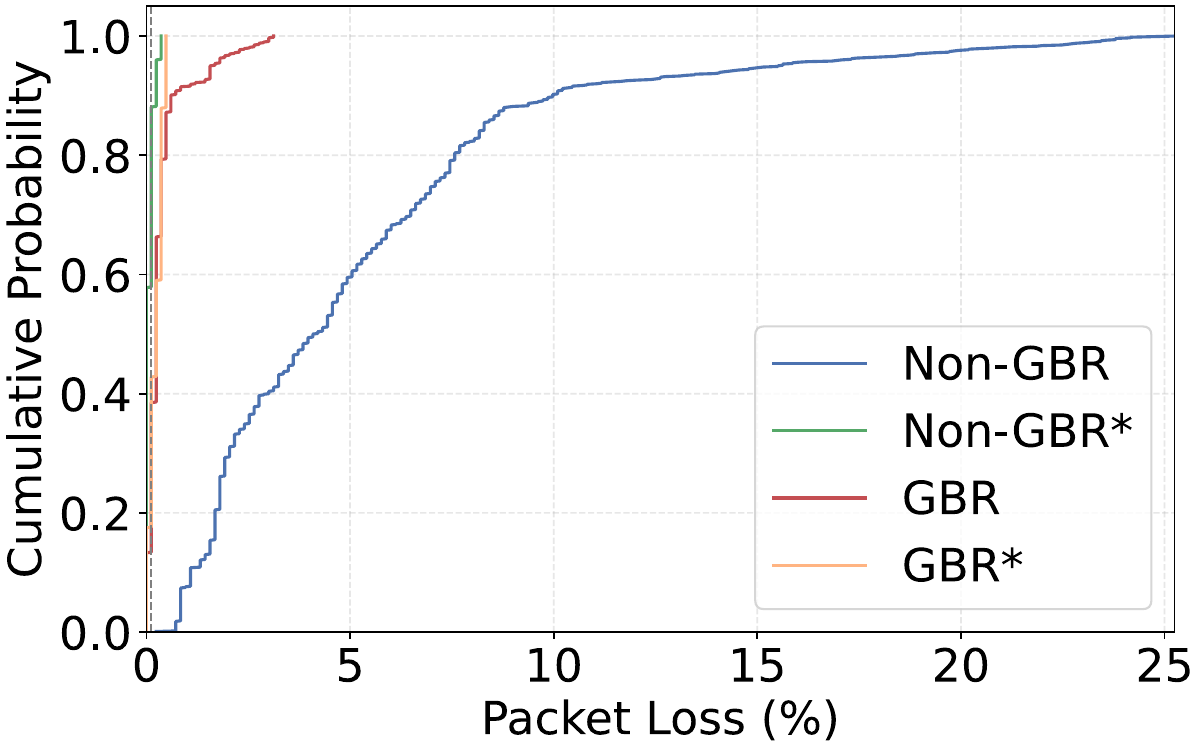}
        \caption{High load}
        \label{fig:cdf-loss-4k}
    \end{subfigure}

    \begin{subfigure}{0.30\linewidth}
        \centering
        \includegraphics[width=\linewidth]{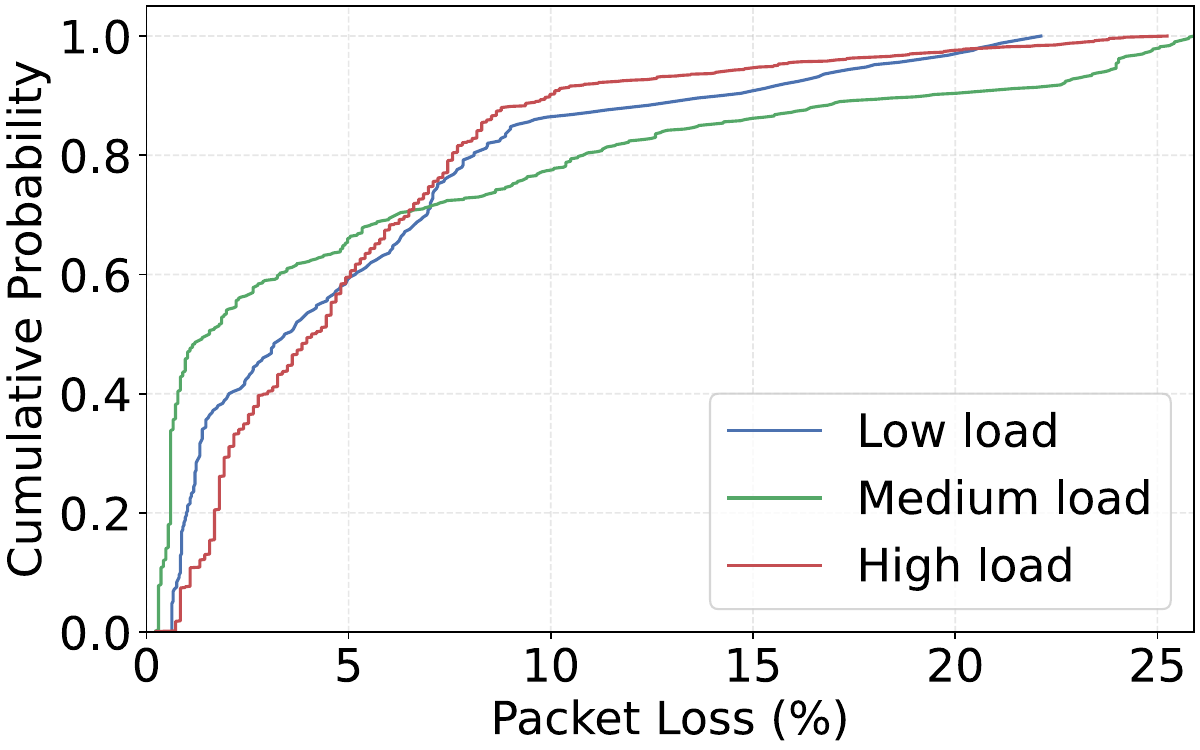}
        \caption{Non-GBR flows}
        \label{fig:cdf-loss-NGBR}
    \end{subfigure}
    \begin{subfigure}{0.30\linewidth}
        \centering
        \includegraphics[width=\linewidth]{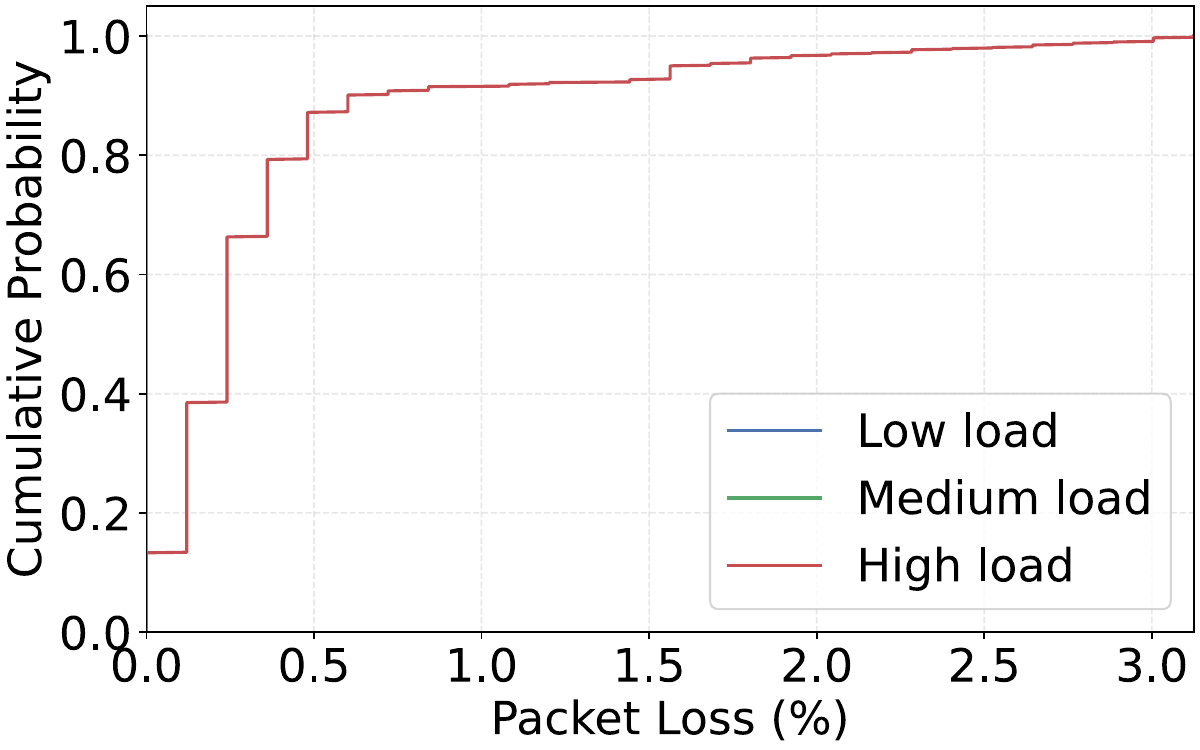}
        \caption{GBR flows}
        \label{fig:cdf-loss-GBR}
    \end{subfigure}

     
    \caption{Throughput and CDF packet loss using full throughput capacity.}
    \label{fig:average_scale}
\end{figure*}

Figure~\ref{fig:average_scale} presents the cumulative distribution functions (CDFs) for throughput and packet loss across all load scenarios. Figures~\ref{fig:cdf-thr-1k} to \ref{fig:cdf-thr-4k} reveal that the throughput for all flows matches their sending rate, except for Non-GBR flows. This demonstrates the ability of our model to isolate throughput degradation to flows lacking a committed rate.
Additionally, it can be observed that throughput remains stable under increasing flow load, showing minimal degradation.

Similarly, the packet loss CDFs in Figures~\ref{fig:cdf-loss-1k} to \ref{fig:cdf-loss-4k} indicate that packet drops are negligible under low and medium loads and only become noticeable for Non-GBR flows at high loads. Further insights are provided in Figures~\ref{fig:cdf-loss-NGBR} and \ref{fig:cdf-loss-GBR}, which show packet loss distributions grouped by resource type, leaving aside the delay-critical, as their packet losses are negligible. The results confirm that packet losses are uncommon for all resource types under low and medium load conditions. Notably, for Non-GBR flows, the loss distribution remains largely unchanged between medium and high load, indicating that loss behavior saturates beyond a certain flow count.

\subsubsection{Throughput compliance with GFBR}~\label{sub:baselineComp}
Here, we aim to evaluate whether our solution meets the per-flow guaranteed bit rate, i.e, GFBR, also referred to as CIR. As a baseline for comparison, we adopt the bandwidth management mechanism proposed in~\cite{related:P4-TINS}, where the authors utilize a single meter to regulate the aggregate throughput of high-level 5G slices (eMBB, mMTC, and uRLLC).  

To analyze per-flow throughput distribution, we configure 2,000 flows within a single GBR-5QI category, all belonging to the same high-level slice. In the traffic pattern, $50\%$ of the flows send traffic at their CIR (4000 Kbps), while the other $50\%$ at their PIR (7000 Kbps). The available bandwidth is sufficient to accommodate the aggregated CIR (i.e., sum of individual CIRs) but not the aggregated PIR. 

\begin{figure}[ht]
    \begin{subfigure}{0.49\linewidth}
        \centering
        \includegraphics[width=0.8\linewidth]{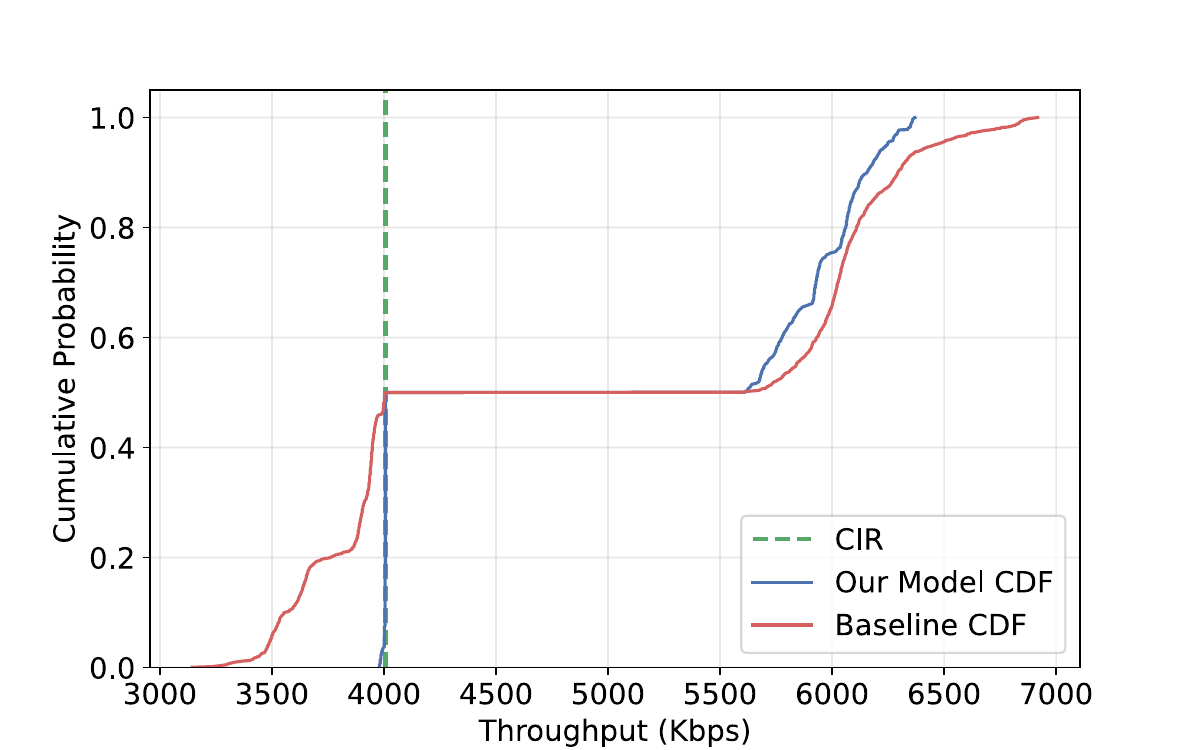}
        \caption{Probability to meet the CIR.}
        \label{fig:throughput-baseline}
    \end{subfigure}
    \begin{subfigure}{0.49\linewidth}
        \centering
        \includegraphics[width=0.8\linewidth]{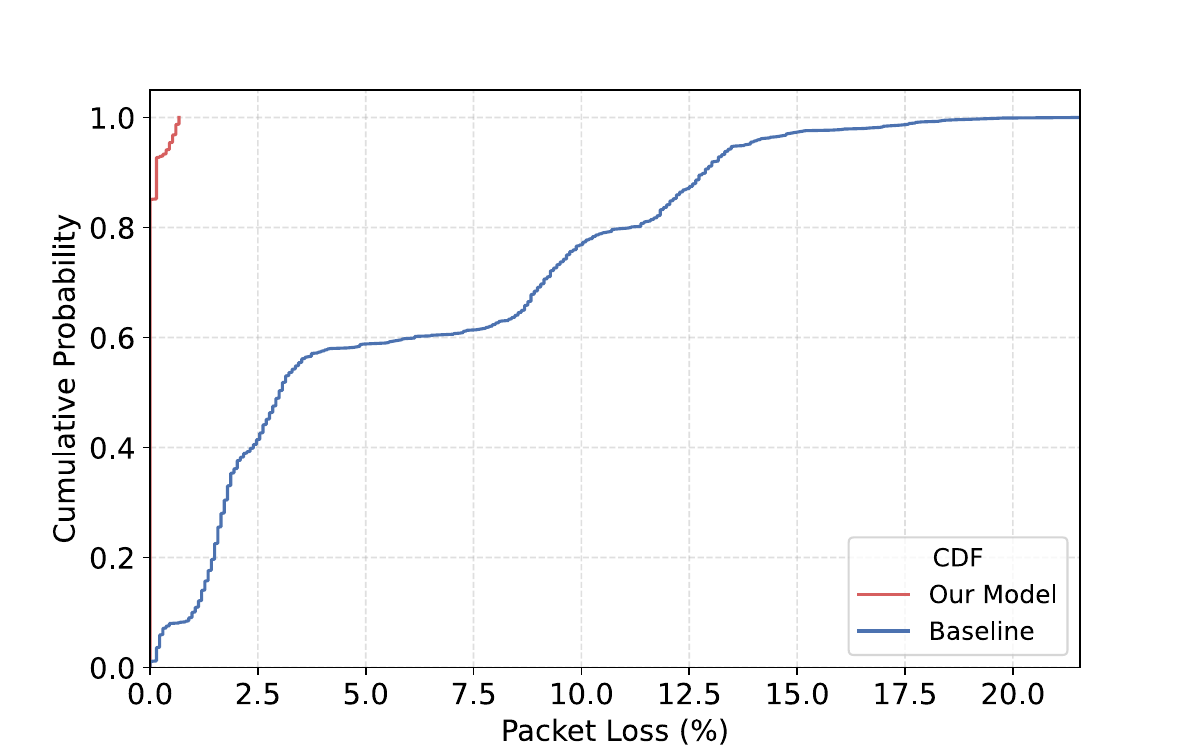}
        \caption{Packet losses for flows sending up to their CIR.}
        \label{fig:losses-baseline}
    \end{subfigure}
    \caption{Per-flow packet losses and throughput comparison between our model and the baseline.}

\end{figure}

Figure~\ref{fig:throughput-baseline} illustrates the CDFs of throughput realized by our model in comparison to the baseline.
Our approach exhibits two key improvements: it consistently guarantees that every flow meets its GFBR (CIR), and it ensures a fairly distribution of the remaining bandwidth.
In contrast, the baseline fails to isolate the impact of flows transmitting at their MFBR (PIR), resulting in throughput degradation for flows constrained to their CIR. This leads to violations of committed bitrate guarantees in approximately 50\% of the flows.

Concerning packet losses, our flow-oriented pipeline ensures near-zero packet loss for flows transmitting at their GFBR while maintaining a fair distribution of losses among flows sending at their MFBR. 
Figure~\ref{fig:losses-baseline} compares the packet loss rate achieved by our model to the baseline. As shown, our solution effectively prevents packet loss for CIR-conforming flows, whereas the baseline exhibits significant loss across all flows. In particular, CIR-conforming flows can experience up to 20\% of packet losses under the baseline model. These results are more favorable to our solution at lower loads (fewer concurrent flows), where the packet losses for these flows are negligible. For page constraints reasons, we provide more details about this in the Appendix~\ref{sec:appendixScale}.


\subsubsection{High-congestion scenario}
This experiment assesses the effectiveness of our prioritization mechanisms under severe congestion, where the aggregate sending rate exceeds the available link capacity. Specifically, we evaluate a medium-flow load scenario in which the total offered load reaches 114\% of the output link capacity. The traffic configuration retains the same 5QI diversity as in previous experiments, with the addition of flows assigned to high-priority level 5QIs. 
To provide more granular insights into per-flow behavior under extreme congestion, additional smaller-scale results are presented in Appendix~\ref{sec:appendixScale}.

Table~\ref{tab:high-con-results} summarizes the experimental setup and the corresponding throughput measurements. In this case, where the aggregate sending rate is 14\% higher than the output link capacity (11.4 Gbps compared to 10 Gbps), packet losses are inevitable. Consistent with prior findings, our model isolates throughput degradation primarily in non-GBR flows. Furthermore, the prioritization mechanism effectively protects prioritized flows, resulting in negligible throughput degradation and near-zero packet loss.

\begin{table}
\caption{Experiment setup and throughput results for a high-congestion scenario.
\dag Aggregated PIR of 2.85 Gbps. \ddag PBS  285 Mb.}
\label{tab:high-con-results}
\scalebox{0.82}{
\begin{tabular}{@{}ccclccccccc@{}}
\toprule
\multirow{2}{*}{5QI} & \multirow{2}{*}{\begin{tabular}[c]{@{}c@{}}Number of \\ Flows\end{tabular}} & \multirow{2}{*}{\begin{tabular}[c]{@{}c@{}}Individual \\ Sending \\ Rate (Kbps)\end{tabular}} & \multicolumn{1}{c}{\multirow{2}{*}{Description}} & \multicolumn{4}{c}{Meters Configuration}                                                                                                                                                                              & \multicolumn{3}{c}{Throughput Results} \\ \cmidrule(l){5-11} 
                     &                                                                             &                                                                                               & \multicolumn{1}{c}{}                             & \begin{tabular}[c]{@{}c@{}}CIR\\ (Kbps)\end{tabular} & \begin{tabular}[c]{@{}c@{}}PIR\\ (Kbps)\end{tabular} & \begin{tabular}[c]{@{}c@{}}CBS\\ (Kb)\end{tabular} & \begin{tabular}[c]{@{}c@{}}PBS\\ (Kb)\end{tabular} & MAX         & MIN         & AVG        \\ \midrule
1                    & 450                                                                         & 6000                                                                                          & Background Non-GBR                               & -                                                    & \dag                                                    & -                                                  & \ddag                                                  & 4954.58     & 2807.59     & 3755.91    \\
2                    & 50                                                                          & 4000                                                                                          & Prioritized Non-GBR                              & -                                                    & \dag                                                    & -                                                  & \ddag                                                  & 4000        & 4000        & 4000       \\
3                    & 450                                                                         & 6000                                                                                          & Background GBR                                   & 2500                                                 & 6000                                                 & 100                                                & 500                                                & 5988.82     & 5455.13     & 5825.46    \\
3                    & 50                                                                          & 1500                                                                                          & Below CIR GBR                                    & 2500                                                 & 6000                                                 & 100                                                & 500                                                & 1500        & 1500        & 1500       \\
4                    & 25                                                                          & 6000                                                                                          & Prioritized GBR                                  & 2500                                                 & 6000                                                 & 100                                                & 500                                                & 6000        & 6000        & 6000       \\
4                    & 25                                                                          & 5000                                                                                          & Prioritized GBR                                  & 2500                                                 & 6000                                                 & 100                                                & 500                                                & 5000        & 5000        & 5000       \\
5                    & 450                                                                         & 6000                                                                                          & Background GBR*                                  & 2500                                                 & 6000                                                 & 100                                                & 500                                                & 5963.48     & 5305.27     & 5801.13    \\
6                    & 450                                                                         & 6000                                                                                          & Background Non-GBR*                              & 2500                                                 & 6000                                                 & 100                                                & 500                                                & 6000        & 5593        & 5833.75    \\ \bottomrule
\end{tabular}
}
\end{table}

\subsubsection{Delay results}

\begin{table*}[!htbp]
  \caption{Maximum delay per queue in milliseconds.}
  \label{tab:delays}
\scalebox{0.8}{
\begin{tabular}{@{}ccccccccccc@{}}
\toprule
Case       & Capacity & Flow Load   & \begin{tabular}[c]{@{}c@{}}$Q_1$ \\ (Non-GBR*)\end{tabular} & \begin{tabular}[c]{@{}c@{}}$Q_2$\\ (GBR*)\end{tabular} & \begin{tabular}[c]{@{}c@{}}$Q_3$\\ (GBR)\end{tabular} & $Q_4$ & \begin{tabular}[c]{@{}c@{}}$Q_5$\\ (Non-GBR*)\end{tabular} & \begin{tabular}[c]{@{}c@{}}$Q_6$\\ (GBR*)\end{tabular} & \begin{tabular}[c]{@{}c@{}}$Q_7$\\ (GBR)\end{tabular} & \begin{tabular}[c]{@{}c@{}}$Q_8$\\ (Non-GBR)\end{tabular} \\ \midrule
Functional & Full     & Low         & 0.119                                                       & 0.134                                                  & 0.21                                                  & -     & 1.536                                                      & 1.639                                                  & 1.751                                                 & 8.224                                                     \\
Functional & Full     & Medium      & 0.237                                                       & 0.249                                                  & 0.81                                                  & -     & 3.77                                                       & 2.812                                                  & 3.844                                                 & 9.21                                                      \\
Functional & Full     & High        & 0.431                                                       & 0.411                                                  & 2.39                                                  & -     & 6.512                                                      & 6.931                                                  & 6.71                                                  & 9.89                                                      \\
Stress     & Full     & 40 flows    & 0.0096                                                      & 0.0095                                                 & 0.15                                                  & -     & -                                                          & -                                                      & -                                                     & 7.13                                                      \\
Stress     & Full     & 400 flows   & 0.058                                                       & 0.059                                                  & 0.37                                                  & -     & -                                                          & -                                                      & -                                                     & 7.66                                                      \\
Stress     & Full     & 4000 flows  & 0.512                                                       & 0.518                                                  & 3.23                                                  & -     & -                                                          & -                                                      & -                                                     & 11.91                                                     \\
Stress     & Full     & 40000 flows & 0.845                                                       & 0.925                                                  & 10.4                                                  & -     & -                                                          & -                                                      & -                                                     & 36.1                                                      \\ \bottomrule
\end{tabular}
}
\end{table*}

To assess packet delay, we measured the time elapsed from the moment the switch receives a packet until it is selected for transmission. In other words, we captured both processing and queueing delays. Although processing delay in conventional fixed-function hardware is typically fixed and minimal, it can become variable and non-negligible in programmable switches, where customized parsing logic and external operations may introduce additional overhead.

As discussed in Section~\ref{sub:PHB}, our model enables the configuration of predictable worst-case queueing delays for delay-critical packets through appropriate buffer sizing and committed rate provisioning. In this section, we evaluate the delay performance of our implementation in two scenarios: a functional deployment representing a realistic use case, and a stress-test scenario designed to assess scalability limits.

Table~\ref{tab:delays} reports the maximum per-queue delay observed in both cases. In the functional scenario, where all flows transmit at rates exceeding their respective CIRs, two distinct delay patterns emerge. 
The first, observed in queues $Q_1$ to $Q_3$, corresponds to packets transmitted within their CIR and reflects the worst-case delay under guaranteed service. The second, observed in $Q_5$ to $Q_7$, captures the delay experienced by packets that exceed their CIR, demonstrating that while traffic beyond the CIR may still be delivered, it incurs additional delay.

These results also reveal a strong correlation between flow load and packet delay. While in \secref{sub:funcEval} we discuss the impact of load on throughput, here we confirm its influence on per-packet delay.

To further analyze this performance degradation, we compare our model against a baseline program implementing a conventional L3 forwarding device.
To ensure a fair comparison, the baseline program replicates our priority queuing policy. 
The experiment uses four distinct 5QIs, one per resource type, with each flow transmitting at its CIR. The aggregate bandwidth per resource type is fixed at $25\%$ of the total output capacity. Initially, each 5QI accommodates 10 flows, and we progressively scale this number by a factor of 10, up to 10,000 flows per 5QI (resulting in 40,000 total flows). This setup maintains a constant aggregate rate per resource type while varying the number of flows, thereby isolating the impact of potential artifacts in our data plane model (e.g., meter execution overhead).

The results of the stress tests using our data plane are also shown in Table~\ref{tab:delays}.
The findings indicate that the maximum packet delay increases as the number of simultaneous flows grows. Furthermore, the delays observed in the baseline program exhibit no significant variability, with a maximum difference of only $0.2$ ms in the queues with the highest variance ($Q_3$ and $Q_8$). This delay is thus attributed to the processing overhead inherent in the specific ASIC hardware. 

Notably, even under the extreme condition of 40,000 concurrent flows, delay-critical packets consistently experience sub-millisecond delay. This demonstrates the robustness of our approach in preserving ultra-low latency guarantees for such traffic, even under adversary conditions.
\section{Discussion}~\label{sec:discussion}
While the proposed P4-programmable data plane for QoS enforcement in the 5GT offers significant improvements in bandwidth guarantees, latency predictability, and traffic prioritization, some challenges and practical considerations must be acknowledged.  

\textit{Hardware Constraints}:  
While programmable data planes offer considerable flexibility and enable rapid innovation, they are still bound by the constraints of underlying hardware. Specifically, the TM components in many commercial programmable switches are not programmable but merely configurable. As a result, the design space for custom packet scheduling algorithms is restricted. This limitation constrains the adoption of more sophisticated scheduling policies, which could further enhance QoS enforcement.  

\textit{Compatibility with Programmable Architectures}:  
Our solution relies on P4 meters for rate enforcement, which limits its applicability to P4 architectures or other programmable hardware that supports this functionality. Although rare, in environments where meters are unavailable, the Policing stage would require an alternative rate-limiting mechanism to achieve similar functionality.  

\textit{Metering Granularity and Accuracy}:
P4 meters utilize token bucket algorithms with predefined granularity, which may result in inaccuracies when enforcing strict per-flow rate guarantees, particularly for highly dynamic traffic patterns.

\textit{Packet Reordering}:  
While assigning packets to different priority queues based on a rate policy ensures throughput guarantees, flows exceeding their committed rate may experience packet reordering. 
This occurs because packets classified into lower-priority queues may encounter variable queueing delays or be dropped, whereas those in higher-priority queues are transmitted with minimal delay. This is a well-known problem in the Internet, for example, and artifacts exist to mitigate packet reordering, including in upper protocol layers.

Addressing these limitations in future work could enhance the scalability and adaptability of the proposed approach, ensuring broader applicability across different network architectures and hardware platforms.

\section{Conclusion}~\label{sec:conclusion}

This paper introduced a QoS-aware programmable data plane model designed to meet the stringent requirements of 5G transport networks. By leveraging per-flow metering, classification, prioritization, and strict priority scheduling, our model ensures predictable latency, bandwidth guarantees, and robust support for differentiated traffic classes, including delay-critical and mission-sensitive flows.

We implemented the proposed architecture using P4 on an Intel Tofino switch and demonstrated its feasibility and performance through extensive experimentation in a real-world testbed. The results confirm that our model achieves per-flow compliance with guaranteed bit rates, isolates degradation to non-guaranteed traffic under congestion, and maintains sub-millisecond latency for high-priority flows, even at scale.

In comparison with existing slice-based resource management solutions, our approach delivers superior fairness, more vigorous SLA enforcement, and more predictable performance under dynamic and traffic conditions.

While certain limitations remain, particularly related to hardware constraints and metering granularity, our work provides a practical and extensible foundation for QoS enforcement in future programmable transport networks. Future directions include exploring adaptive scheduling algorithms, cross-layer integration with control-plane analytics, and extending the model to support end-to-end QoS guarantees.  




\begin{thebibliography}{40}


\ifx \showCODEN    \undefined \def \showCODEN     #1{\unskip}     \fi
\ifx \showISBNx    \undefined \def \showISBNx     #1{\unskip}     \fi
\ifx \showISBNxiii \undefined \def \showISBNxiii  #1{\unskip}     \fi
\ifx \showISSN     \undefined \def \showISSN      #1{\unskip}     \fi
\ifx \showLCCN     \undefined \def \showLCCN      #1{\unskip}     \fi
\ifx \shownote     \undefined \def \shownote      #1{#1}          \fi
\ifx \showarticletitle \undefined \def \showarticletitle #1{#1}   \fi
\ifx \showURL      \undefined \def \showURL       {\relax}        \fi
\providecommand\bibfield[2]{#2}
\providecommand\bibinfo[2]{#2}
\providecommand\natexlab[1]{#1}
\providecommand\showeprint[2][]{arXiv:#2}

\bibitem[3GPP(2024a)]%
        {3gpp:slicing:TS28.530}
\bibfield{author}{\bibinfo{person}{3GPP}.} \bibinfo{year}{2024}\natexlab{a}.
\newblock \bibinfo{booktitle}{\emph{{3rd Generation Partnership Project; Technical Specification Group Services and System Aspects; Management and orchestration; Concepts, use cases and requirements (Release 18)}}}.
\newblock \bibinfo{type}{{T}echnical {R}eport} TS 28.530 V18.1.0 (2024-09). \bibinfo{institution}{3GPP}.
\newblock


\bibitem[3GPP(2024b)]%
        {3gpp:Priority:TS22.261}
\bibfield{author}{\bibinfo{person}{3GPP}.} \bibinfo{year}{2024}\natexlab{b}.
\newblock \bibinfo{booktitle}{\emph{{3rd Generation Partnership Project; Technical Specification Group Services and System Aspects; Service requirements for the 5G system; Stage 1 (Release 20)}}}.
\newblock \bibinfo{type}{{T}echnical {R}eport} TS 22.261 V20.0.0 (2024-09). \bibinfo{institution}{3GPP}.
\newblock


\bibitem[Alizadeh et~al\mbox{.}(2013)]%
        {pFabric}
\bibfield{author}{\bibinfo{person}{Mohammad Alizadeh}, \bibinfo{person}{Shuang Yang}, \bibinfo{person}{Milad Sharif}, \bibinfo{person}{Sachin Katti}, \bibinfo{person}{Nick McKeown}, \bibinfo{person}{Balaji Prabhakar}, {and} \bibinfo{person}{Scott Shenker}.} \bibinfo{year}{2013}\natexlab{}.
\newblock \showarticletitle{{pFabric: minimal near-optimal datacenter transport}}.
\newblock \bibinfo{journal}{\emph{SIGCOMM Comput. Commun. Rev.}} \bibinfo{volume}{43}, \bibinfo{number}{4} (\bibinfo{date}{Aug.} \bibinfo{year}{2013}), \bibinfo{pages}{435–446}.
\newblock
\showISSN{0146-4833}
\href{https://doi.org/10.1145/2534169.2486031}{doi:\nolinkurl{10.1145/2534169.2486031}}


\bibitem[{Anonymous}(2025)]%
        {code-repo}
\bibfield{author}{\bibinfo{person}{{Anonymous}}.} \bibinfo{year}{2025}\natexlab{}.
\newblock \bibinfo{title}{GitHub Repository}.
\newblock
\newblock
\shownote{URL and details omitted for double-blind reviewing}.


\bibitem[Balasingam et~al\mbox{.}(2025)]%
        {RAN1}
\bibfield{author}{\bibinfo{person}{Arjun Balasingam}, \bibinfo{person}{Manikanta Kotaru}, {and} \bibinfo{person}{Paramvir Bahl}.} \bibinfo{year}{2025}\natexlab{}.
\newblock \showarticletitle{{Application-level service assurance with 5G RAN slicing}}. In \bibinfo{booktitle}{\emph{Proceedings of the 21st USENIX Symposium on Networked Systems Design and Implementation}} (Santa Clara, CA, USA) \emph{(\bibinfo{series}{NSDI'24})}. \bibinfo{publisher}{USENIX Association}, \bibinfo{address}{USA}, Article \bibinfo{articleno}{47}, \bibinfo{numpages}{17}~pages.
\newblock
\showISBNx{978-1-939133-39-7}


\bibitem[Bosshart et~al\mbox{.}(2014)]%
        {P4}
\bibfield{author}{\bibinfo{person}{Pat Bosshart}, \bibinfo{person}{Dan Daly}, \bibinfo{person}{Glen Gibb}, \bibinfo{person}{Martin Izzard}, \bibinfo{person}{Nick McKeown}, \bibinfo{person}{Jennifer Rexford}, \bibinfo{person}{Cole Schlesinger}, \bibinfo{person}{Dan Talayco}, \bibinfo{person}{Amin Vahdat}, \bibinfo{person}{George Varghese}, {and} \bibinfo{person}{David Walker}.} \bibinfo{year}{2014}\natexlab{}.
\newblock \showarticletitle{{P4: programming protocol-independent packet processors}}.
\newblock \bibinfo{journal}{\emph{SIGCOMM Comput. Commun. Rev.}} \bibinfo{volume}{44}, \bibinfo{number}{3} (\bibinfo{date}{July} \bibinfo{year}{2014}), \bibinfo{pages}{87–95}.
\newblock
\showISSN{0146-4833}
\href{https://doi.org/10.1145/2656877.2656890}{doi:\nolinkurl{10.1145/2656877.2656890}}


\bibitem[Bosshart et~al\mbox{.}(2013)]%
        {progammability1}
\bibfield{author}{\bibinfo{person}{Pat Bosshart}, \bibinfo{person}{Glen Gibb}, \bibinfo{person}{Hun-Seok Kim}, \bibinfo{person}{George Varghese}, \bibinfo{person}{Nick McKeown}, \bibinfo{person}{Martin Izzard}, \bibinfo{person}{Fernando Mujica}, {and} \bibinfo{person}{Mark Horowitz}.} \bibinfo{year}{2013}\natexlab{}.
\newblock \showarticletitle{{Forwarding metamorphosis: fast programmable match-action processing in hardware for SDN}}. In \bibinfo{booktitle}{\emph{Proceedings of the ACM SIGCOMM 2013 Conference on SIGCOMM}} (Hong Kong, China) \emph{(\bibinfo{series}{SIGCOMM '13})}. \bibinfo{publisher}{Association for Computing Machinery}, \bibinfo{address}{New York, NY, USA}, \bibinfo{pages}{99–110}.
\newblock
\showISBNx{9781450320566}
\href{https://doi.org/10.1145/2486001.2486011}{doi:\nolinkurl{10.1145/2486001.2486011}}


\bibitem[Chen et~al\mbox{.}(2023)]%
        {RAN15}
\bibfield{author}{\bibinfo{person}{Yongzhou Chen}, \bibinfo{person}{Ruihao Yao}, \bibinfo{person}{Haitham Hassanieh}, {and} \bibinfo{person}{Radhika Mittal}.} \bibinfo{year}{2023}\natexlab{}.
\newblock \showarticletitle{{Channel-Aware 5G RAN Slicing with Customizable Schedulers}}. In \bibinfo{booktitle}{\emph{20th USENIX Symposium on Networked Systems Design and Implementation (NSDI 23)}}. \bibinfo{publisher}{USENIX Association}, \bibinfo{address}{Boston, MA}, \bibinfo{pages}{1767--1782}.
\newblock
\showISBNx{978-1-939133-33-5}
\urldef\tempurl%
\url{https://www.usenix.org/conference/nsdi23/presentation/chen-yongzhou}
\showURL{%
\tempurl}


\bibitem[Chen et~al\mbox{.}(2022)]%
        {related:P4-TINS}
\bibfield{author}{\bibinfo{person}{Yan-Wei Chen}, \bibinfo{person}{Chi-Yu Li}, \bibinfo{person}{Chien-Chao Tseng}, {and} \bibinfo{person}{Min-Zhi Hu}.} \bibinfo{year}{2022}\natexlab{}.
\newblock \showarticletitle{{P4-TINS: P4-Driven Traffic Isolation for Network Slicing With Bandwidth Guarantee and Management}}.
\newblock \bibinfo{journal}{\emph{IEEE Transactions on Network and Service Management}} \bibinfo{volume}{19}, \bibinfo{number}{3} (\bibinfo{year}{2022}), \bibinfo{pages}{3290--3303}.
\newblock
\href{https://doi.org/10.1109/TNSM.2022.3159232}{doi:\nolinkurl{10.1109/TNSM.2022.3159232}}


\bibitem[Chen et~al\mbox{.}(2024)]%
        {progammability3}
\bibfield{author}{\bibinfo{person}{Zhikang Chen}, \bibinfo{person}{Yong Feng}, \bibinfo{person}{Shuxin Liu}, \bibinfo{person}{Haoyu Song}, \bibinfo{person}{Hanyi Zhou}, \bibinfo{person}{Tong Yun}, \bibinfo{person}{Wenquan Xu}, \bibinfo{person}{Tian Pan}, {and} \bibinfo{person}{Bin Liu}.} \bibinfo{year}{2024}\natexlab{}.
\newblock \showarticletitle{{OptimusPrime: Unleash Dataplane Programmability through a Transformable Architecture}}. In \bibinfo{booktitle}{\emph{Proceedings of the ACM SIGCOMM 2024 Conference}} (Sydney, NSW, Australia) \emph{(\bibinfo{series}{ACM SIGCOMM '24})}. \bibinfo{publisher}{Association for Computing Machinery}, \bibinfo{address}{New York, NY, USA}, \bibinfo{pages}{904–920}.
\newblock
\showISBNx{9798400706141}
\href{https://doi.org/10.1145/3651890.3672214}{doi:\nolinkurl{10.1145/3651890.3672214}}


\bibitem[Chunduri et~al\mbox{.}(2025)]%
        {ietf-dmm-tn-aware-mobility}
\bibfield{author}{\bibinfo{person}{Uma Chunduri}, \bibinfo{person}{John Kaippallimalil}, \bibinfo{person}{Sridhar Bhaskaran}, \bibinfo{person}{Jeff Tantsura}, {and} \bibinfo{person}{Luis~M. Contreras}.} \bibinfo{year}{2025}\natexlab{}.
\newblock \bibinfo{booktitle}{\emph{Mobility-aware Transport Network Slicing for 5G}}.
\newblock \bibinfo{type}{Internet-Draft} draft-ietf-dmm-tn-aware-mobility-19. \bibinfo{institution}{IETF Secretariat}.
\newblock


\bibitem[Corporation(2021)]%
        {IntelTNA}
\bibfield{author}{\bibinfo{person}{Intel Corporation}.} \bibinfo{year}{2021}\natexlab{}.
\newblock \bibinfo{title}{{P416 Intel{\textregistered} Tofino{\texttrademark} Native Architecture}}.
\newblock \bibinfo{howpublished}{Online}.
\newblock
\urldef\tempurl%
\url{https://raw.githubusercontent.com/barefootnetworks/Open-Tofino/master/PUBLIC_Tofino-Native-Arch.pdf}
\showURL{%
\tempurl}
\newblock
\shownote{\url{https://raw.githubusercontent.com/barefootnetworks/Open-Tofino/master/PUBLIC_Tofino-Native-Arch.pdf}. Accessed: 2025-03-28}.


\bibitem[Corporation(2025)]%
        {IntelTofino}
\bibfield{author}{\bibinfo{person}{Intel Corporation}.} \bibinfo{year}{2025}\natexlab{}.
\newblock \bibinfo{title}{{Intel Tofino Series}}.
\newblock
\newblock
\shownote{\url{https://www.intel.com/content/www/us/en/products/details/network-io/intelligent-fabric-processors/tofino.html}. Accessed: 2025-03-25}.


\bibitem[Elbediwy et~al\mbox{.}(2024)]%
        {related:DRPIFO}
\bibfield{author}{\bibinfo{person}{Mostafa Elbediwy}, \bibinfo{person}{Bill Pontikakis}, \bibinfo{person}{Alireza Ghaffari}, \bibinfo{person}{Jean-Pierre David}, {and} \bibinfo{person}{Yvon Savaria}.} \bibinfo{year}{2024}\natexlab{}.
\newblock \showarticletitle{{DR-PIFO: A Dynamic Ranking Packet Scheduler Using a Push-In-First-Out Queue}}.
\newblock \bibinfo{journal}{\emph{IEEE Transactions on Network and Service Management}} \bibinfo{volume}{21}, \bibinfo{number}{1} (\bibinfo{year}{2024}), \bibinfo{pages}{355--371}.
\newblock
\href{https://doi.org/10.1109/TNSM.2023.3304894}{doi:\nolinkurl{10.1109/TNSM.2023.3304894}}


\bibitem[Foukas et~al\mbox{.}(2017)]%
        {RAN24}
\bibfield{author}{\bibinfo{person}{Xenofon Foukas}, \bibinfo{person}{Mahesh~K. Marina}, {and} \bibinfo{person}{Kimon Kontovasilis}.} \bibinfo{year}{2017}\natexlab{}.
\newblock \showarticletitle{{Orion: RAN Slicing for a Flexible and Cost-Effective Multi-Service Mobile Network Architecture}}. In \bibinfo{booktitle}{\emph{Proceedings of the 23rd Annual International Conference on Mobile Computing and Networking}} (Snowbird, Utah, USA) \emph{(\bibinfo{series}{MobiCom '17})}. \bibinfo{publisher}{Association for Computing Machinery}, \bibinfo{address}{New York, NY, USA}, \bibinfo{pages}{127–140}.
\newblock
\showISBNx{9781450349161}
\href{https://doi.org/10.1145/3117811.3117831}{doi:\nolinkurl{10.1145/3117811.3117831}}


\bibitem[Gao et~al\mbox{.}(2024)]%
        {related:sifter}
\bibfield{author}{\bibinfo{person}{Peixuan Gao}, \bibinfo{person}{Anthony Dalleggio}, \bibinfo{person}{Jiajin Liu}, \bibinfo{person}{Chen Peng}, \bibinfo{person}{Yang Xu}, {and} \bibinfo{person}{H.~Jonathan Chao}.} \bibinfo{year}{2024}\natexlab{}.
\newblock \showarticletitle{{Sifter: an inversion-free and large-capacity programmable packet scheduler}}. In \bibinfo{booktitle}{\emph{Proceedings of the 21st USENIX Symposium on Networked Systems Design and Implementation}} (Santa Clara, CA, USA) \emph{(\bibinfo{series}{NSDI'24})}. \bibinfo{publisher}{USENIX Association}, \bibinfo{address}{USA}, Article \bibinfo{articleno}{5}, \bibinfo{numpages}{20}~pages.
\newblock
\showISBNx{978-1-939133-39-7}


\bibitem[Geng et~al\mbox{.}(2023)]%
        {ietf-teas-5g-network-slice-application}
\bibfield{author}{\bibinfo{person}{Xuesong Geng}, \bibinfo{person}{Luis~M. Contreras}, \bibinfo{person}{Reza Rokui}, \bibinfo{person}{Jie Dong}, {and} \bibinfo{person}{Ivan Bykov}.} \bibinfo{year}{2023}\natexlab{}.
\newblock \bibinfo{booktitle}{\emph{IETF Network Slice Application in 3GPP 5G End-to-End Network Slice}}.
\newblock \bibinfo{type}{Internet-Draft} draft-ietf-teas-5g-network-slice-application-01. \bibinfo{institution}{IETF Secretariat}.
\newblock


\bibitem[Harkous et~al\mbox{.}(2021)]%
        {related:virtual-queues}
\bibfield{author}{\bibinfo{person}{Hasanin Harkous}, \bibinfo{person}{Chrysa Papagianni}, \bibinfo{person}{Koen De~Schepper}, \bibinfo{person}{Michael Jarschel}, \bibinfo{person}{Marinos Dimolianis}, {and} \bibinfo{person}{Rastin Pries}.} \bibinfo{year}{2021}\natexlab{}.
\newblock \showarticletitle{{Virtual Queues for P4: A Poor Man’s Programmable Traffic Manager}}.
\newblock \bibinfo{journal}{\emph{IEEE Transactions on Network and Service Management}} \bibinfo{volume}{18}, \bibinfo{number}{3} (\bibinfo{year}{2021}), \bibinfo{pages}{2860--2872}.
\newblock
\href{https://doi.org/10.1109/TNSM.2021.3077051}{doi:\nolinkurl{10.1109/TNSM.2021.3077051}}


\bibitem[Hasan et~al\mbox{.}(2023)]%
        {bwg9}
\bibfield{author}{\bibinfo{person}{Shaddi Hasan}, \bibinfo{person}{Amar Padmanabhan}, \bibinfo{person}{Bruce Davie}, \bibinfo{person}{Jennifer Rexford}, \bibinfo{person}{Ulas Kozat}, \bibinfo{person}{Hunter Gatewood}, \bibinfo{person}{Shruti Sanadhya}, \bibinfo{person}{Nick Yurchenko}, \bibinfo{person}{Tariq Al-Khasib}, \bibinfo{person}{Oriol Batalla}, \bibinfo{person}{Marie Bremner}, \bibinfo{person}{Andrei Lee}, \bibinfo{person}{Evgeniy Makeev}, \bibinfo{person}{Scott Moeller}, \bibinfo{person}{Alex Rodriguez}, \bibinfo{person}{Pravin Shelar}, \bibinfo{person}{Karthik Subraveti}, \bibinfo{person}{Sudarshan Kandi}, \bibinfo{person}{Alejandro Xoconostle}, \bibinfo{person}{Praveen Kumar~Ramakrishnan}, {and} \bibinfo{person}{Xiaochen Tian}.} \bibinfo{year}{2023}\natexlab{}.
\newblock \showarticletitle{{Building Flexible, Low-Cost Wireless Access Networks with Magma}}.
\newblock \bibinfo{journal}{\emph{GetMobile: Mobile Comp. and Comm.}} \bibinfo{volume}{27}, \bibinfo{number}{3} (\bibinfo{date}{Nov.} \bibinfo{year}{2023}), \bibinfo{pages}{40–47}.
\newblock
\showISSN{2375-0529}
\href{https://doi.org/10.1145/3631588.3631599}{doi:\nolinkurl{10.1145/3631588.3631599}}


\bibitem[Hauser et~al\mbox{.}(2022)]%
        {related:slicingP4}
\bibfield{author}{\bibinfo{person}{Eric Hauser}, \bibinfo{person}{Manuel Simon}, \bibinfo{person}{Henning Stubbe}, \bibinfo{person}{Sebastian Gallenm\"{u}ller}, {and} \bibinfo{person}{Georg Carle}.} \bibinfo{year}{2022}\natexlab{}.
\newblock \showarticletitle{{Slicing networks with P4 hardware and software targets}}. In \bibinfo{booktitle}{\emph{Proceedings of the ACM SIGCOMM Workshop on 5G and Beyond Network Measurements, Modeling, and Use Cases}} (Amsterdam, Netherlands) \emph{(\bibinfo{series}{5G-MeMU '22})}. \bibinfo{publisher}{Association for Computing Machinery}, \bibinfo{address}{New York, NY, USA}, \bibinfo{pages}{36–42}.
\newblock
\showISBNx{9781450393935}
\href{https://doi.org/10.1145/3538394.3546043}{doi:\nolinkurl{10.1145/3538394.3546043}}


\bibitem[Hauser et~al\mbox{.}(2023)]%
        {survey-p4-5g}
\bibfield{author}{\bibinfo{person}{Frederik Hauser}, \bibinfo{person}{Marco Häberle}, \bibinfo{person}{Daniel Merling}, \bibinfo{person}{Steffen Lindner}, \bibinfo{person}{Vladimir Gurevich}, \bibinfo{person}{Florian Zeiger}, \bibinfo{person}{Reinhard Frank}, {and} \bibinfo{person}{Michael Menth}.} \bibinfo{year}{2023}\natexlab{}.
\newblock \showarticletitle{{A survey on data plane programming with P4: Fundamentals, advances, and applied research}}.
\newblock \bibinfo{journal}{\emph{Journal of Network and Computer Applications}}  \bibinfo{volume}{212} (\bibinfo{year}{2023}), \bibinfo{pages}{103561}.
\newblock
\showISSN{1084-8045}
\href{https://doi.org/10.1016/j.jnca.2022.103561}{doi:\nolinkurl{10.1016/j.jnca.2022.103561}}


\bibitem[Heinanen and Guerin(1999a)]%
        {srtcm-RFC2697}
\bibfield{author}{\bibinfo{person}{J. Heinanen} {and} \bibinfo{person}{R. Guerin}.} \bibinfo{year}{1999}\natexlab{a}.
\newblock \bibinfo{booktitle}{\emph{{A Single Rate Three Color Marker}}}.
\newblock \bibinfo{type}{RFC} 2697. \bibinfo{institution}{RFC Editor}.
\newblock
\showISSN{2070-1721}


\bibitem[Heinanen and Guerin(1999b)]%
        {trtcm-RFC2698}
\bibfield{author}{\bibinfo{person}{J. Heinanen} {and} \bibinfo{person}{R. Guerin}.} \bibinfo{year}{1999}\natexlab{b}.
\newblock \bibinfo{booktitle}{\emph{{A Two Rate Three Color Marker}}}.
\newblock \bibinfo{type}{RFC} 2698. \bibinfo{institution}{RFC Editor}.
\newblock
\showISSN{2070-1721}


\bibitem[Kazmi et~al\mbox{.}(2019)]%
        {book:slicing:kazmi2019}
\bibfield{author}{\bibinfo{person}{SM~Ahsan Kazmi}, \bibinfo{person}{Latif~U Khan}, \bibinfo{person}{Nguyen~H Tran}, {and} \bibinfo{person}{Choong~Seon Hong}.} \bibinfo{year}{2019}\natexlab{}.
\newblock \bibinfo{booktitle}{\emph{{Network slicing for 5G and beyond networks}}}. Vol.~\bibinfo{volume}{1}.
\newblock \bibinfo{publisher}{Springer}.
\newblock


\bibitem[Kfoury et~al\mbox{.}(2021)]%
        {related:p4-survey-traffic-policy}
\bibfield{author}{\bibinfo{person}{Elie~F. Kfoury}, \bibinfo{person}{Jorge Crichigno}, {and} \bibinfo{person}{Elias Bou-Harb}.} \bibinfo{year}{2021}\natexlab{}.
\newblock \showarticletitle{{An Exhaustive Survey on P4 Programmable Data Plane Switches: Taxonomy, Applications, Challenges, and Future Trends}}.
\newblock \bibinfo{journal}{\emph{IEEE Access}}  \bibinfo{volume}{9} (\bibinfo{year}{2021}), \bibinfo{pages}{87094--87155}.
\newblock
\href{https://doi.org/10.1109/ACCESS.2021.3086704}{doi:\nolinkurl{10.1109/ACCESS.2021.3086704}}


\bibitem[Kokku et~al\mbox{.}(2012)]%
        {RAN41}
\bibfield{author}{\bibinfo{person}{Ravi Kokku}, \bibinfo{person}{Rajesh Mahindra}, \bibinfo{person}{Honghai Zhang}, {and} \bibinfo{person}{Sampath Rangarajan}.} \bibinfo{year}{2012}\natexlab{}.
\newblock \showarticletitle{{NVS: a substrate for virtualizing wireless resources in cellular networks}}.
\newblock \bibinfo{journal}{\emph{IEEE/ACM Trans. Netw.}} \bibinfo{volume}{20}, \bibinfo{number}{5} (\bibinfo{date}{Oct.} \bibinfo{year}{2012}), \bibinfo{pages}{1333–1346}.
\newblock
\showISSN{1063-6692}
\href{https://doi.org/10.1109/TNET.2011.2179063}{doi:\nolinkurl{10.1109/TNET.2011.2179063}}


\bibitem[Li et~al\mbox{.}(2019)]%
        {Core46}
\bibfield{author}{\bibinfo{person}{Xin Li}, \bibinfo{person}{Chengcheng Guo}, \bibinfo{person}{Lav Gupta}, {and} \bibinfo{person}{Raj Jain}.} \bibinfo{year}{2019}\natexlab{}.
\newblock \showarticletitle{{Efficient and Secure 5G Core Network Slice Provisioning Based on VIKOR Approach}}.
\newblock \bibinfo{journal}{\emph{IEEE Access}}  \bibinfo{volume}{7} (\bibinfo{year}{2019}), \bibinfo{pages}{150517--150529}.
\newblock
\href{https://doi.org/10.1109/ACCESS.2019.2947454}{doi:\nolinkurl{10.1109/ACCESS.2019.2947454}}


\bibitem[MacDavid et~al\mbox{.}(2024)]%
        {bw-slicing}
\bibfield{author}{\bibinfo{person}{Robert MacDavid}, \bibinfo{person}{Xiaoqi Chen}, {and} \bibinfo{person}{Jennifer Rexford}.} \bibinfo{year}{2024}\natexlab{}.
\newblock \showarticletitle{{Scalable Real-Time Bandwidth Fairness in Switches}}.
\newblock \bibinfo{journal}{\emph{IEEE/ACM Transactions on Networking}} \bibinfo{volume}{32}, \bibinfo{number}{2} (\bibinfo{year}{2024}), \bibinfo{pages}{1423--1434}.
\newblock
\href{https://doi.org/10.1109/TNET.2023.3317172}{doi:\nolinkurl{10.1109/TNET.2023.3317172}}


\bibitem[Netronome(2024)]%
        {netronomeSmartnics}
\bibfield{author}{\bibinfo{person}{Netronome}.} \bibinfo{year}{2024}\natexlab{}.
\newblock \bibinfo{title}{{Agilio SmartNICs}}.
\newblock
\urldef\tempurl%
\url{https://netronome.com/agilio-smartnics/}
\showURL{%
\tempurl}
\newblock
\shownote{\url{https://netronome.com/agilio-smartnics/}. Accessed: 2025-04-03}.


\bibitem[Pittalà et~al\mbox{.}(2024)]%
        {related:dataplaneprogrammability}
\bibfield{author}{\bibinfo{person}{Gaetano~Francesco Pittalà}, \bibinfo{person}{Lorenzo Rinieri}, \bibinfo{person}{Amir {Al Sadi}}, \bibinfo{person}{Gianluca Davoli}, \bibinfo{person}{Andrea Melis}, \bibinfo{person}{Marco Prandini}, {and} \bibinfo{person}{Walter Cerroni}.} \bibinfo{year}{2024}\natexlab{}.
\newblock \showarticletitle{{Leveraging Data Plane Programmability to enhance service orchestration at the edge: A focus on industrial security}}.
\newblock \bibinfo{journal}{\emph{Computer Networks}}  \bibinfo{volume}{246} (\bibinfo{year}{2024}), \bibinfo{pages}{110397}.
\newblock
\showISSN{1389-1286}
\href{https://doi.org/10.1016/j.comnet.2024.110397}{doi:\nolinkurl{10.1016/j.comnet.2024.110397}}


\bibitem[Ricart-Sanchez et~al\mbox{.}(2020)]%
        {related:QoS-Aware}
\bibfield{author}{\bibinfo{person}{Ruben Ricart-Sanchez}, \bibinfo{person}{Pedro Malagon}, \bibinfo{person}{Antonio Matencio-Escolar}, \bibinfo{person}{Jose~M. Alcaraz~Calero}, {and} \bibinfo{person}{Qi Wang}.} \bibinfo{year}{2020}\natexlab{}.
\newblock \showarticletitle{{Toward hardware-accelerated QoS-aware 5G network slicing based on data plane programmability}}.
\newblock \bibinfo{journal}{\emph{Transactions on Emerging Telecommunications Technologies}} \bibinfo{volume}{31}, \bibinfo{number}{1} (\bibinfo{year}{2020}), \bibinfo{pages}{e3726}.
\newblock
\href{https://doi.org/10.1002/ett.3726}{doi:\nolinkurl{10.1002/ett.3726}}
\newblock
\shownote{e3726 ett.3726}.


\bibitem[Sharma et~al\mbox{.}(2020)]%
        {pscheduler2}
\bibfield{author}{\bibinfo{person}{Naveen~Kr. Sharma}, \bibinfo{person}{Chenxingyu Zhao}, \bibinfo{person}{Ming Liu}, \bibinfo{person}{Pravein~G Kannan}, \bibinfo{person}{Changhoon Kim}, \bibinfo{person}{Arvind Krishnamurthy}, {and} \bibinfo{person}{Anirudh Sivaraman}.} \bibinfo{year}{2020}\natexlab{}.
\newblock \showarticletitle{{Programmable calendar queues for high-speed packet scheduling}}. In \bibinfo{booktitle}{\emph{Proceedings of the 17th Usenix Conference on Networked Systems Design and Implementation}} (Santa Clara, CA, USA) \emph{(\bibinfo{series}{NSDI'20})}. \bibinfo{publisher}{USENIX Association}, \bibinfo{address}{USA}, \bibinfo{pages}{685–700}.
\newblock
\showISBNx{9781939133137}


\bibitem[Shreedhar and Varghese(1995)]%
        {paper:DWRR}
\bibfield{author}{\bibinfo{person}{M. Shreedhar} {and} \bibinfo{person}{George Varghese}.} \bibinfo{year}{1995}\natexlab{}.
\newblock \showarticletitle{{Efficient fair queueing using deficit round robin}}. In \bibinfo{booktitle}{\emph{Proceedings of the Conference on Applications, Technologies, Architectures, and Protocols for Computer Communication}} (Cambridge, Massachusetts, USA) \emph{(\bibinfo{series}{SIGCOMM '95})}. \bibinfo{publisher}{Association for Computing Machinery}, \bibinfo{address}{New York, NY, USA}, \bibinfo{pages}{231–242}.
\newblock
\showISBNx{0897917111}
\href{https://doi.org/10.1145/217382.217453}{doi:\nolinkurl{10.1145/217382.217453}}


\bibitem[Wang et~al\mbox{.}(2020)]%
        {related:TCP-friendly}
\bibfield{author}{\bibinfo{person}{Shie-Yuan Wang}, \bibinfo{person}{Hsien-Wen Hu}, {and} \bibinfo{person}{Yi-Bing Lin}.} \bibinfo{year}{2020}\natexlab{}.
\newblock \showarticletitle{{Design and Implementation of TCP-Friendly Meters in P4 Switches}}.
\newblock \bibinfo{journal}{\emph{IEEE/ACM Transactions on Networking}} \bibinfo{volume}{28}, \bibinfo{number}{4} (\bibinfo{year}{2020}), \bibinfo{pages}{1885--1898}.
\newblock
\href{https://doi.org/10.1109/TNET.2020.3002074}{doi:\nolinkurl{10.1109/TNET.2020.3002074}}


\bibitem[Xing et~al\mbox{.}(2022)]%
        {progammability2}
\bibfield{author}{\bibinfo{person}{Jiarong Xing}, \bibinfo{person}{Kuo-Feng Hsu}, \bibinfo{person}{Matty Kadosh}, \bibinfo{person}{Alan Lo}, \bibinfo{person}{Yonatan Piasetzky}, \bibinfo{person}{Arvind Krishnamurthy}, {and} \bibinfo{person}{Ang Chen}.} \bibinfo{year}{2022}\natexlab{}.
\newblock \showarticletitle{{Runtime Programmable Switches}}. In \bibinfo{booktitle}{\emph{19th USENIX Symposium on Networked Systems Design and Implementation (NSDI 22)}}. \bibinfo{publisher}{USENIX Association}, \bibinfo{address}{Renton, WA}, \bibinfo{pages}{651--665}.
\newblock
\showISBNx{978-1-939133-27-4}
\urldef\tempurl%
\url{https://www.usenix.org/conference/nsdi22/presentation/xing}
\showURL{%
\tempurl}


\bibitem[Xing et~al\mbox{.}(2023)]%
        {progammability-smartNICs}
\bibfield{author}{\bibinfo{person}{Jiarong Xing}, \bibinfo{person}{Yiming Qiu}, \bibinfo{person}{Kuo-Feng Hsu}, \bibinfo{person}{Songyuan Sui}, \bibinfo{person}{Khalid Manaa}, \bibinfo{person}{Omer Shabtai}, \bibinfo{person}{Yonatan Piasetzky}, \bibinfo{person}{Matty Kadosh}, \bibinfo{person}{Arvind Krishnamurthy}, \bibinfo{person}{T.~S.~Eugene Ng}, {and} \bibinfo{person}{Ang Chen}.} \bibinfo{year}{2023}\natexlab{}.
\newblock \showarticletitle{{Unleashing SmartNIC Packet Processing Performance in P4}}. In \bibinfo{booktitle}{\emph{Proceedings of the ACM SIGCOMM 2023 Conference}} (New York, NY, USA) \emph{(\bibinfo{series}{ACM SIGCOMM '23})}. \bibinfo{publisher}{Association for Computing Machinery}, \bibinfo{address}{New York, NY, USA}, \bibinfo{pages}{1028–1042}.
\newblock
\showISBNx{9798400702365}
\href{https://doi.org/10.1145/3603269.3604882}{doi:\nolinkurl{10.1145/3603269.3604882}}


\bibitem[Yan et~al\mbox{.}(2024)]%
        {related:example-number-flows}
\bibfield{author}{\bibinfo{person}{Jinzhu Yan}, \bibinfo{person}{Haotian Xu}, \bibinfo{person}{Zhuotao Liu}, \bibinfo{person}{Qi Li}, \bibinfo{person}{Ke Xu}, \bibinfo{person}{Mingwei Xu}, {and} \bibinfo{person}{Jianping Wu}.} \bibinfo{year}{2024}\natexlab{}.
\newblock \showarticletitle{{Brain-on-switch: towards advanced intelligent network data plane via NN-driven traffic analysis at line-speed}}. In \bibinfo{booktitle}{\emph{Proceedings of the 21st USENIX Symposium on Networked Systems Design and Implementation}} (Santa Clara, CA, USA) \emph{(\bibinfo{series}{NSDI'24})}. \bibinfo{publisher}{USENIX Association}, \bibinfo{address}{USA}, Article \bibinfo{articleno}{24}, \bibinfo{numpages}{22}~pages.
\newblock
\showISBNx{978-1-939133-39-7}


\bibitem[Yang et~al\mbox{.}(2024)]%
        {progammability-runtime}
\bibfield{author}{\bibinfo{person}{Yifan Yang}, \bibinfo{person}{Lin He}, \bibinfo{person}{Jiasheng Zhou}, \bibinfo{person}{Xiaoyi Shi}, \bibinfo{person}{Jiamin Cao}, {and} \bibinfo{person}{Ying Liu}.} \bibinfo{year}{2024}\natexlab{}.
\newblock \showarticletitle{{P4runpro: Enabling Runtime Programmability for RMT Programmable Switches}}. In \bibinfo{booktitle}{\emph{Proceedings of the ACM SIGCOMM 2024 Conference}} (Sydney, NSW, Australia) \emph{(\bibinfo{series}{ACM SIGCOMM '24})}. \bibinfo{publisher}{Association for Computing Machinery}, \bibinfo{address}{New York, NY, USA}, \bibinfo{pages}{921–937}.
\newblock
\showISBNx{9798400706141}
\href{https://doi.org/10.1145/3651890.3672230}{doi:\nolinkurl{10.1145/3651890.3672230}}


\bibitem[Ye et~al\mbox{.}(2018)]%
        {Core68}
\bibfield{author}{\bibinfo{person}{Qiang Ye}, \bibinfo{person}{Junling Li}, \bibinfo{person}{Kaige Qu}, \bibinfo{person}{Weihua Zhuang}, \bibinfo{person}{Xuemin~Sherman Shen}, {and} \bibinfo{person}{Xu Li}.} \bibinfo{year}{2018}\natexlab{}.
\newblock \showarticletitle{{End-to-End Quality of Service in 5G Networks: Examining the Effectiveness of a Network Slicing Framework}}.
\newblock \bibinfo{journal}{\emph{IEEE Vehicular Technology Magazine}} \bibinfo{volume}{13}, \bibinfo{number}{2} (\bibinfo{year}{2018}), \bibinfo{pages}{65--74}.
\newblock
\href{https://doi.org/10.1109/MVT.2018.2809473}{doi:\nolinkurl{10.1109/MVT.2018.2809473}}


\bibitem[Yoo and Chen(2021)]%
        {hash-collision}
\bibfield{author}{\bibinfo{person}{Sophia Yoo} {and} \bibinfo{person}{Xiaoqi Chen}.} \bibinfo{year}{2021}\natexlab{}.
\newblock \showarticletitle{{Secure Keyed Hashing on Programmable Switches}}. In \bibinfo{booktitle}{\emph{Proceedings of the ACM SIGCOMM 2021 Workshop on Secure Programmable Network INfrastructure}} (Virtual Event, USA) \emph{(\bibinfo{series}{SPIN '21})}. \bibinfo{publisher}{Association for Computing Machinery}, \bibinfo{address}{New York, NY, USA}, \bibinfo{pages}{16–22}.
\newblock
\showISBNx{9781450386371}
\href{https://doi.org/10.1145/3472873.3472881}{doi:\nolinkurl{10.1145/3472873.3472881}}


\end{thebibliography}


\newpage
\appendix
\section{Glossary of Terms}

\begin{table}[h!]
\caption{List of Terms}
\scalebox{0.7}{
\begin{tabular}{@{}ll@{}}
\toprule
\multicolumn{1}{c}{\textbf{Term}} & \multicolumn{1}{c}{\textbf{Definition}} \\ \midrule
5GC                               & 5G Core Network                         \\
5GT                               & 5G Transport Network                    \\
5QI                               & 5G QoS Identifiers                      \\
ARP                               & Allocation and Retention Priority                      \\
CBS                               & Committed Burst Size                    \\
CIR                               & Committed Information Rate              \\
CU                                & Central Unit                            \\
DU                                & Distributed Unit                        \\
GBR                               & Guarantee Bit Rate                      \\
GBR*                              & Delay-Critical Guarantee Bit Rate       \\
GFBR                              & Guaranteed Flow Bit Rate                \\
MDBV                              & Maximum Data Burst Volume               \\
MEC                               & Multi-access Edge Computing             \\
MFBR                              & Maximum Flow Bit Rate                \\
Non-GBR                           & Non-Guarantee Bit Rate                  \\
Non-GBR*                          & Delay-Critical Non-Guarantee Bit Rate   \\
PBS                               & Peak Burst Size                         \\
PER                               & Packet Error Rate                     \\
PDB                               & Packet Delay Budget                     \\
PIR                               & Peak Information Rate                   \\
QoS                               & Quality of Service                      \\
RAN                               & Radio Access Network                    \\
RU                                & Radio Unit                              \\
TEID                              & Tunnel Endpoint Identifier              \\
TNA                               & Tofino Native Architecture              \\
UPF                               & User Plane Function                     \\ \bottomrule
\end{tabular}
}
\end{table}

\section{Additional results at small scales}\label{sec:appendixScale}
In the previous sections, we conducted experiments involving realistic amounts of flows and per-flow sending rates. Here, we will present other experiments with a limited number of high-rate flows to precisely characterize per-flow behaviors in our switch implementation.

\subsection{Functional evaluation}
As in~\secref{sub:funcEval}, here we conduct a series of tests with aggregated sending rates reaching up to 100\% of the link capacity. The scenarios include varying numbers of flows per 5QI, each with different sending rates. Flows were configured to send traffic under all possible conditions: below the CIR, up to the CIR, up to the PIR, and beyond the PIR.

When the aggregate sending rate remains below 100\% of the link capacity, the switch successfully serves all flows without packet losses while maintaining a maximum delay of less than 1 ms. The only exception applies to flows exceeding their PIR, as they violate the throughput policy. Consequently, a portion of their packets is discarded to regulate their sending rate, ensuring it remains within the PIR limit and preventing network congestion.

\begin{figure}[ht]
  \centering
  \includegraphics[width=0.6\linewidth]{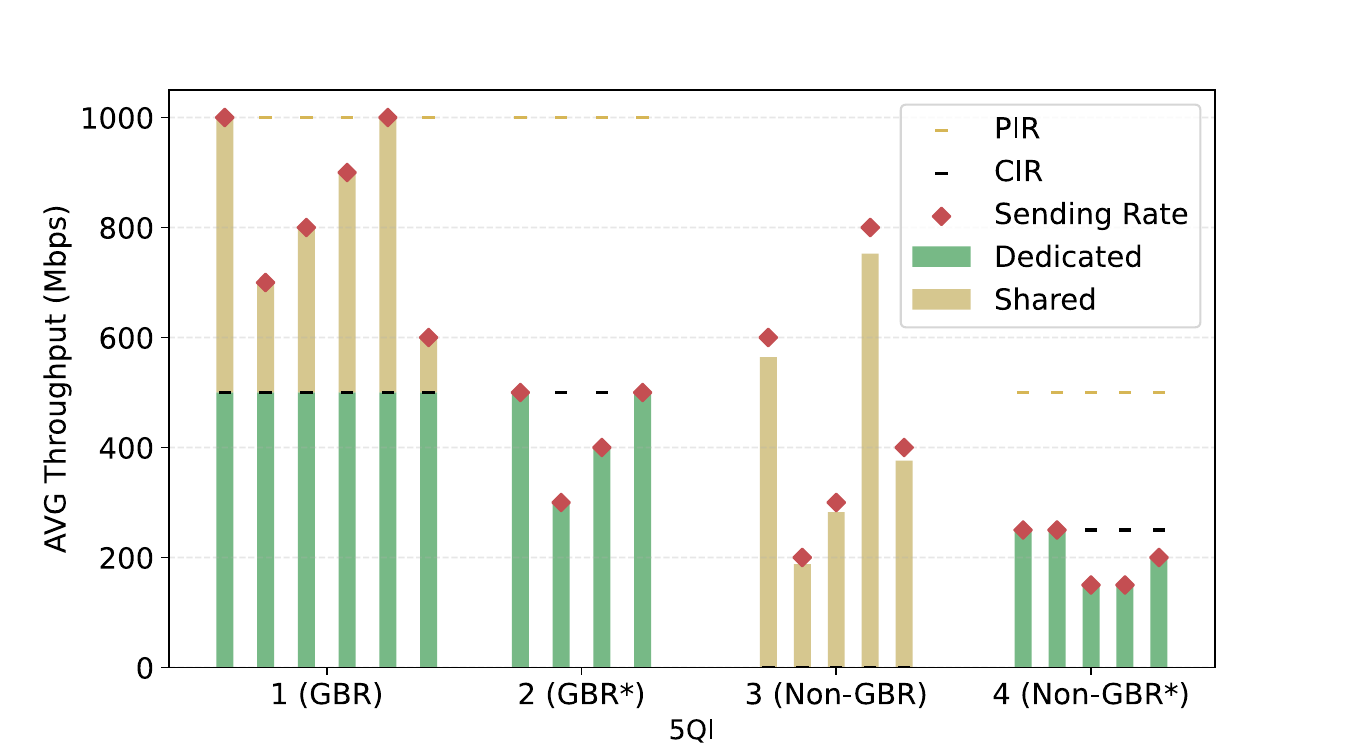}
  \caption{Per-flow throughput results for the functional evaluation. Each bar represents a flow within the 5QI. }
  \label{fig:throughput-1-4-small}
\end{figure}

The relevant scenario occurs when the aggregate sending rate fully compromises the link capacity. Figure~\ref{fig:throughput-1-4-small} presents the throughput results for a scenario with this aggregate sending rate and where GBR flows transmit between their CIR and PIR, GBR* and Non-GBR* up to their CIR, and Non-GBR flows collectively match their PIR. Bars are colored by packet classification: green for bit rate guarantee and yellow for best-effort service.

The results indicate that the throughput for all flows matches their sending rate, except for Non-GBR flows. Throughput variance is not visible in the figure, as it is zero for GBR, Non-GBR*, and GBR* flows, while for Non-GBR flows, it reaches up to $2.1$ Mbps.

Although GBR flows transmit above their CIR, and consequently, some of their packets are enqueued in shared resources, these flows still achieve their intended sending rates and have no packet losses. This demonstrates the per-resource-type traffic isolation capability of our design. For comparison, the aggregated rate of packets absorbed by the low-priority queues (yellow bars) is $2.3$ Gbps for Non-GBR flows and $2$ Gbps for GBR flows.

Another noteworthy observation is the fair distribution of packet losses among Non-GBR flows. In this scenario, where flows have varying sending rates, the measured average packet loss rate is 5.9\%, with a variance of 0.001\%.

\subsection{Throughput accuracy}
Here, we aim to expand the throughput evaluation against the baseline shown in~\secref{sub:baselineComp}.
To analyze per-flow throughput distribution, we configure multiple flows within a single GBR-5QI category, all belonging to the same high-level slice. The traffic pattern includes flows transmitting at their CIR and others at their PIR, while ensuring that the available output bandwidth is always sufficient to accommodate the aggregated CIR but not the aggregated PIR.  

\begin{figure}[ht]
    \begin{subfigure}{0.49\linewidth}
        \centering
        \includegraphics[width=\linewidth]{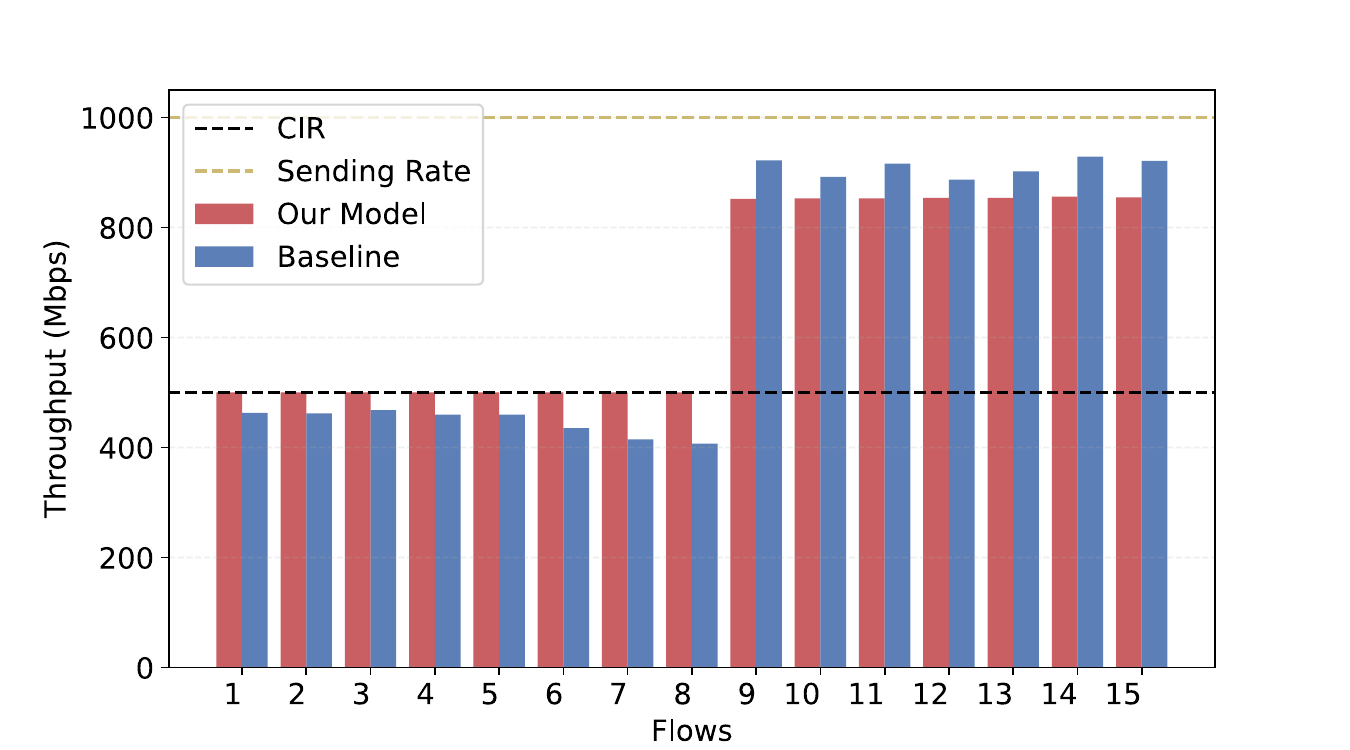}
        \caption{Throughput.}
        \label{fig:throughput-baseline-small}
    \end{subfigure}
    \begin{subfigure}{0.49\linewidth}
        \centering
        \includegraphics[width=\linewidth]{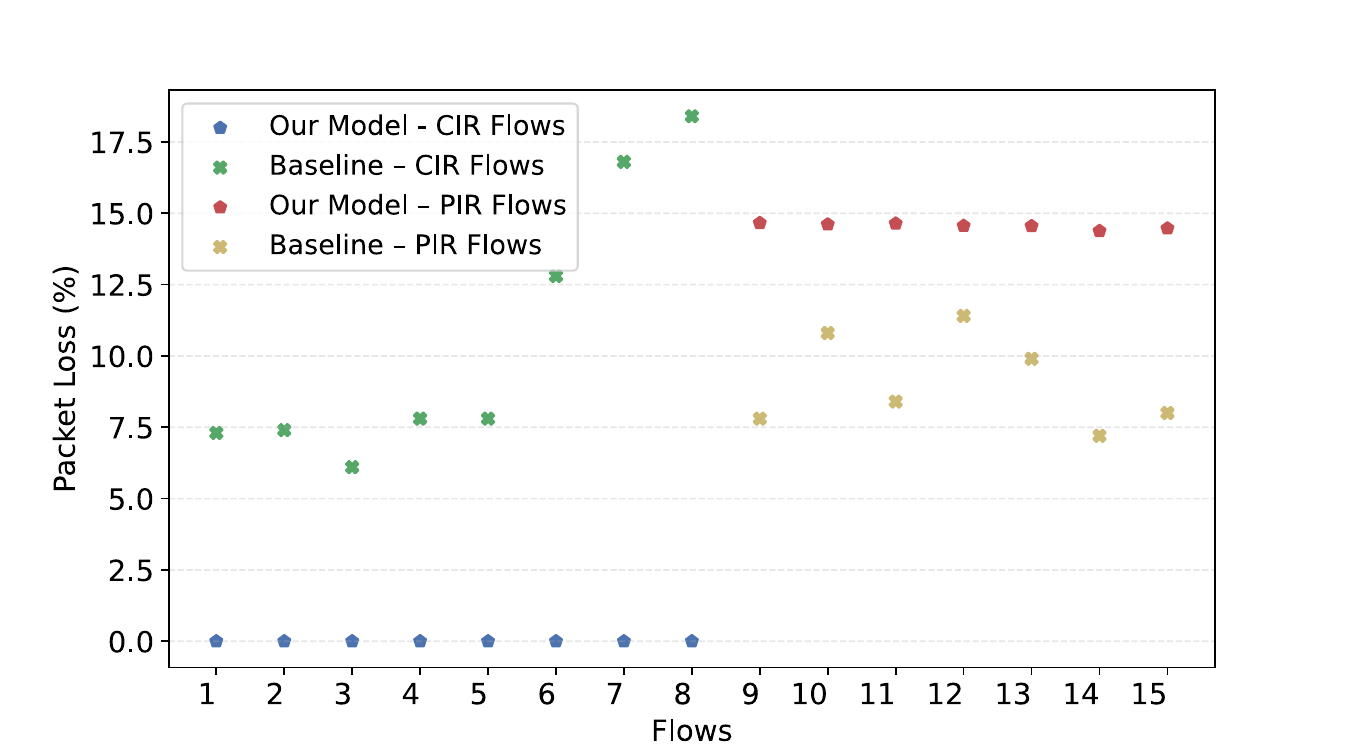}
        \caption{Packet losses.}
        \label{fig:losses-baseline-small}
    \end{subfigure}
    \caption{Per-flow packet losses and throughput comparison between our model and the baseline.}

\end{figure}

Figure~\ref{fig:throughput-baseline-small} illustrates the throughput results for each flow in our model compared to the baseline. In the figure, flows 1 to 8 send traffic at their CIR rate (500 Mbps), while the others send at their PIR (1000 Mbps).

Our model exhibits two key improvements: first, it consistently guarantees that every flow meets its CIR, and secondly, it ensures a fair distribution of the remaining bandwidth. 
In contrast, although the baseline effectively enforces the aggregate CIR rate, it fails to isolate the impact of flows transmitting at the PIR, affecting the throughput of flows sending up to their CIRs, thereby compromising the bit rate guarantees.

Concerning packet losses, our per-flow treatment ensures zero packet loss for flows transmitting at their CIR while maintaining a fair distribution of losses among flows sending at their PIR. Figure~\ref{fig:losses-baseline-small} presents a comparison of packet loss percentages per flow. As observed, in our switch, packet losses occur exclusively in flows exceeding their CIR, and these losses are evenly distributed. Conversely, in the baseline switch, packet losses affect all flows without a discernible distribution pattern.

\subsection{High-congestion scenario}
In this scenario, we configured aggregated sending rates reaching up to 110\% of the link capacity, while maintaining the diversity of 5QIs and their respective individual sending rates. Additionally, we added some flows belonging to 5QIs with a high-priority value.

Although we performed several experiments varying the number of flows and adjusting the CIR and PIR proportionally, we selected an experiment with a low number of flows per 5QI and large CIR and PIR values for better visualization. Table~\ref{tab:configs-small} shows the configuration used for this experiment.

Figure~\ref{fig:throughput10-small} presents the measured throughput for this configuration, revealing the following key observations: 
\begin{inparaenum}[i)]
\item all flows reach their CIR,
\item delay-critical flows achieve their intended sending rate,
\item prioritized flows successfully reach their sending rate, and
\item throughput loss due to congestion is equitably distributed among flows of the same resource type, ensuring isolation.
\end{inparaenum}
Furthermore, throughput variance is negligible, with a maximum observed deviation of 3.4 Mbps, which is imperceptible in the figure.

In summary, this experiment demonstrates that, even under congestion, our model effectively prioritizes critical traffic (e.g., from mission-critical applications), significantly enhancing its performance while maintaining resource-type isolation, low delays, and GBR requirements.


\begin{table}[h!]
  \centering
  \begin{minipage}[t]{0.48\columnwidth}
    \centering
    \caption{High-Congestion scenario setup. The PIR specified under 5QI 3 is aggregated among all Non-GBR flows.}
  \label{tab:configs-small}
    \scalebox{0.8}{
    \begin{tabular}{@{}clcccc@{}}
\toprule
5QI & \multicolumn{1}{c}{\begin{tabular}[c]{@{}c@{}}Resource\\ Type\end{tabular}} & Flows & \begin{tabular}[c]{@{}c@{}}CIR\\ (Mbps)\end{tabular} & \begin{tabular}[c]{@{}c@{}}PIR\\ (Mbps)\end{tabular} & Priority \\ \midrule
0   & GBR                                                                         & 6     & 500                                                  & 1000                                                 & 50       \\
1   & GBR                                                                         & 1     & 500                                                  & 1000                                                 & 5        \\
2   & GBR*                                                                        & 4     & 500                                                  & 1000                                                 & 25       \\
3   & Non-GBR                                                                     & 3     & -                                                    & 2500                                                 & 50       \\
4   & Non-GBR                                                                     & 1     &                                                      &                                                      & 5        \\
5   & Non-GBR*                                                                    & 4     & 250                                                  & 500                                                  & 60       \\ \bottomrule
\end{tabular}
}
  \end{minipage}%
  \hfill
  \begin{minipage}[t]{0.49\columnwidth}
    \centering
    \caption{Per-flow throughput results for the high-congestion scenario.}
    \label{fig:throughput10-small}
    \centering
    \includegraphics[width=\linewidth]{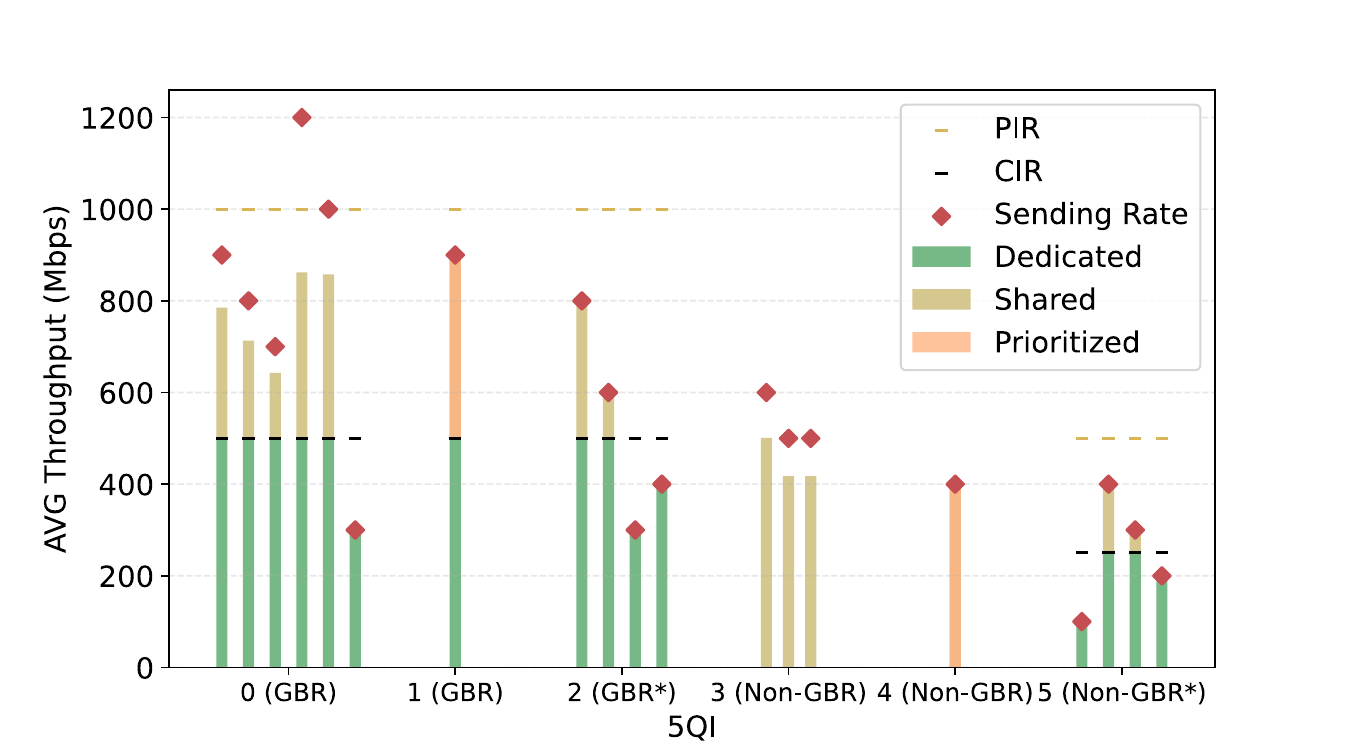}
     
  \end{minipage}
\end{table}


\end{document}